# Efficient variational synthesis of quantum circuits with coherent multi-start optimization

Nikita A. Nemkov[1,2], Evgeniy O. Kiktenko[1,2], Ilia A. Luchnikov[1,2], and Aleksey K. Fedorov[1,2]

[1]Russian Quantum Center, Skolkovo, Moscow 143026, Russia
[2]National University of Science and Technology "MISIS", Moscow 119049, Russia

**We consider the problem of the variational quantum circuit synthesis into a gate set consisting of the CNOT gate and arbitrary single-qubit (1q) gates, with the primary objective being the minimization of the CNOT count. First, we note that along with the discrete architecture search, suffering from the combinatorial explosion of complexity, optimization over 1q gates can also be a crucial roadblock due to the omnipresence of local minimums (well known in the context of variational quantum algorithms but apparently underappreciated in the context of the variational compiling). Taking the issue seriously, we make an extensive search over the initial conditions an essential part of our approach. Another key idea we propose is to use parametrized two-qubit (2q) controlled phase gates, which can interpolate between the identity gate and the CNOT gate, and allow a continuous relaxation of the discrete architecture search, which can be executed jointly with the optimization over 1q gates. This coherent optimization of the architecture together with 1q gates appears to work surprisingly well in practice, sometimes even outperforming optimization over 1q gates alone (for fixed optimal architectures). As illustrative examples and applications we derive 8 CNOT and T depth 3 decomposition of the 3q Toffoli gate on the nearest-neighbor topology, rediscover best known decompositions of the 4q Toffoli gate on all 4q topologies including a 1 CNOT gate improvement on the star-shaped topology, and propose decomposition of the 5q Toffoli gate on the nearest-neighbor topology with 48 CNOT gates. We also benchmark the performance of our approach on a number of 5q quantum circuits from the ibm_qx_mapping database, showing that it is highly competitive with the existing software. The algorithm developed in this work is available as a Python package CPFlow .**

Nikita A. Nemkov: nnemkov@gmail.com
Aleksey K. Fedorov: akf@rqc.ru

## Contents

## 1 Introduction

While many quantum algorithms, such as integer factoring [1], unstructured search [2], or linear equation solvers [3], promise game-changing speedups over classical, the current state of the quantum computing technology does not yet allow for a decisive demonstration with useful applications, although it might be on the verge (see e.g. a recent review [4]). There are plenty of factors limiting the performance of the current generation of quantum devices, such as initialization and readout errors, loss of coherence over time, and errors in gate operations. In the current NISQ era [5], most gate-based quantum protocols are constructed out of single-qubit (1q) and two-qubit (2q) gates with the latter being significantly more error-prone across all leading platforms (at the same time, the problem of realizing multi-qubit gates is also under active study, see e.g. Ref. [6]). Hence, minimization of the 2q gate count is one of the key objectives that can improve performance of the near-term algorithms. On the other hand, in the fault-tolerant future a different type of resource, e.g. the T gate count, is likely to be the most expensive.

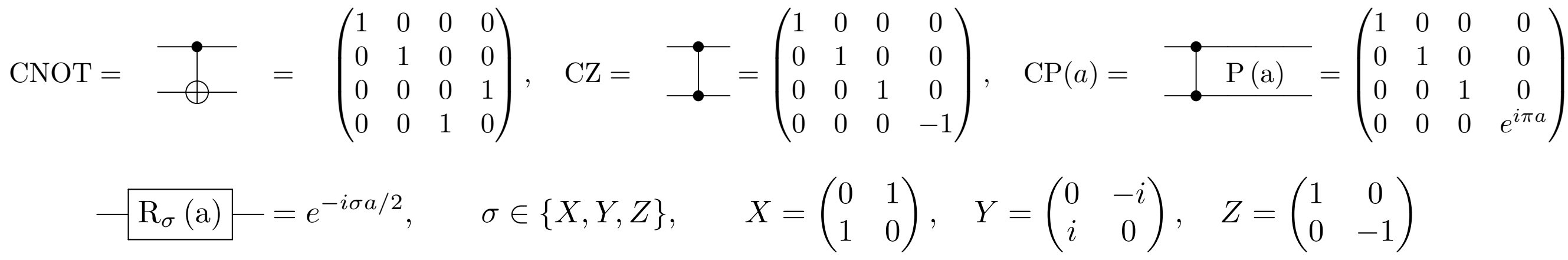

Figure 1: Diagrammatic notation and matrix representation for gates used in this work. When drawing large circuits we will often abbreviate rotation gates $R_\sigma(a)$ by $\sigma(a)$ to lighten the notation.

At the high level, quantum algorithms are usually described using primitives, such as large multi-controlled gates or quantum Fourier transform, that are not directly accessible on the current devices. The standard compiling routines [7, 8] include decomposing algorithmic primitives into native gates (which is always possible [9]), routing stage to comply with the possible connectivity restrictions of the target chip, and local simplifications of the resulting circuits (see e.g. here for a benchmark comparison of different frameworks [10]). Using special data structures such as graphs, tensor networks, ZX diagrams, decision diagrams and others, allows carrying out the compilation process without the need to simulate any part of the circuit. This makes these techniques extremely scalable, allowing to compile and optimize circuits with hundreds of qubits. The downside is that the resulting decompositions may be significantly less efficient than possible. A complementary strategy is to work directly with the unitary matrix of the circuit, thus eliminating any potential redundancies or inefficiencies in the original gate-based description. This is only feasible for small scale circuits, as the size of the state space and the corresponding unitary matrices scales exponentially with the number of qubits. In fact, even for a few-qubit circuits there may be other limiting factors such as circuit complexity, as we emphasize in this work. Although there could be applications of direct unitary synthesis to enhancing the performance of the NISQ algorithms, we expect the use of highly optimized small scale circuits as building blocks of large scale algorithms to be the most promising possibility.

The unitary synthesis problem amounts to finding the most efficient circuits optimizing a given objective function. Typical applications include maximizing fidelity with respect to the target unitary (compilation) or maximizing the overlap with the target state (state preparation), but more general problems can be considered. We study the problem of the variational synthesis into the gate set consisting of a single 2q gate (CZ or CNOT ) and arbitrary 1q gates. The primary optimization objective is the amount of CNOT gates, although indirectly we also address CNOT depth and even T count and T depth. It is natural to divide the problem into two parts:

(i) Discrete optimization or architecture search, looking for best placements of 2q gates.

(ii) Continuous optimization of 1q gate parameters for a given architecture.

The difficulty associated with the architecture search has combinatorial origin and is manifest. The difficulty of the continuous optimization is however also essential, as is known in the context of variational quantum algorithms, but apparently underappreciated in the context of the variational compiling.

After fixing our notation and giving a brief introduction in Sec. 2, we zoom in the issue of the continuous optimization in Sec. 3. Results of this analysis may be of independent interest. Another central ingredient in our approach is to use parametrized 2q gates as a means to relax the discrete architecture search to another continuous optimization, that can be performed simultaneously and coherently with the optimization over 1q gates. We introduce the CPFlow algorithm in Sec. 4. In Sec. 5 we first illustrate all central features of CPFlow using the 3q Toffoli gate as an example, and then go beyond this toy case to synthesize efficient (and likely novel) decompositions of the 4q and 5q Toffoli gates on constrained topologies. In Sec. 6, we provide further benchmarks showcasing that CPFlow is very efficient in compilation of small scale circuits with moderate complexity, but also outline its limitations. Sec. 7 concludes with the summary and outlook.

There are three main contributions we present in this study.

(i) Exposing the problem of local minimums in the loss landscapes of variational compiling problems as a crucial yet underappreciated challenge.

(ii) Developing a new heuristic algorithm for a simul-

taneous search over the architectures and single-qubit angles and demonstrating that for small-scale quantum circuits of intermediate complexity it can produce optimal or nearly optimal decompositions.

(iii) Proposing a simple post-processing step, that often allows refining approximate decompositions into exact.

Observation (i) and technique (iii) are not specific to our core algorithm and are likely to be of wider interest.

Optimization of quantum circuits implementing various tasks, from simulation to combinatorial optimization, is critical for successful applications. Some approaches are informed by the structure of the problem, e.g. the symmetry of the dynamics to be simulated [11, 12, 13, 14]. Others seek for compilation layouts robust to noise [15, 16, 17, 18] or even attempt to redefine the basis gate set based on the chip design [19]. Our work is better characterized as numerical circuit synthesis.

The idea to use computer assisted search and numerical optimization for circuit synthesis goes back a long way [20] and continues to the present day with advances due to both algorithm design and growth of raw computational power. Possible frameworks include purely discrete search over a finite gate set [21, 22], a natural separation into discrete architecture search and continuous optimization [23, 24], adaptive circuit synthesis [25, 26, 27, 28], techniques such as genetic algorithms [29, 30] and machine learning [31, 32], and a hybrid approach with part of the architecture search outsourced to a version of continuous optimization [33, 34]. The scheme developed in Ref. [34] is in many respects similar to the one proposed in this paper, and similarly to our work, was originally motivated by the impressive success of variational compiling of random unitaries [35, 36, 37]. Random unitaries can be thought of as the circuits with maximal complexity. In this work, we address circuits of intermediate complexity, which are a much more challenging target.

# 2 Background and notation

## 2.1 Quantum circuits and quantum gates

Quantum circuits are usually drawn as diagrams similar to Fig. 4. Horizontal wires correspond to qubits, boxes and vertical connections to quantum gates. Each quantum gate can be though of as a unitary matrix. The ones we use in this paper are summarized in Fig. 1. An important detail is that CZ and CNOT gates are equivalent up to a conjugation by the 1q Hadamard gate, and hence completely equivalent for the purpose of the variational compiling that we consider. We often mention CNOT gates in more general discussions, as is more standard, but refer to CZ gates when the technical details are important. The state space associated to $n$ qubits has dimension $2^n$, the unitary matrices acting on this space have dimension $2^n \times 2^n$. The unitary matrix of a quantum circuit can be constructed by an appropriate tensor product of all the gates involved. The overall complex phase of the unitary matrix is inconsequential and can be fixed in any convenient way. The space of quantum circuits on $n$ qubits is hence equivalent to the special unitary group $SU(2^n)$. A natural measure of fidelity between two unitary matrices is the distance induced from the Hilbert-Schmidt norm

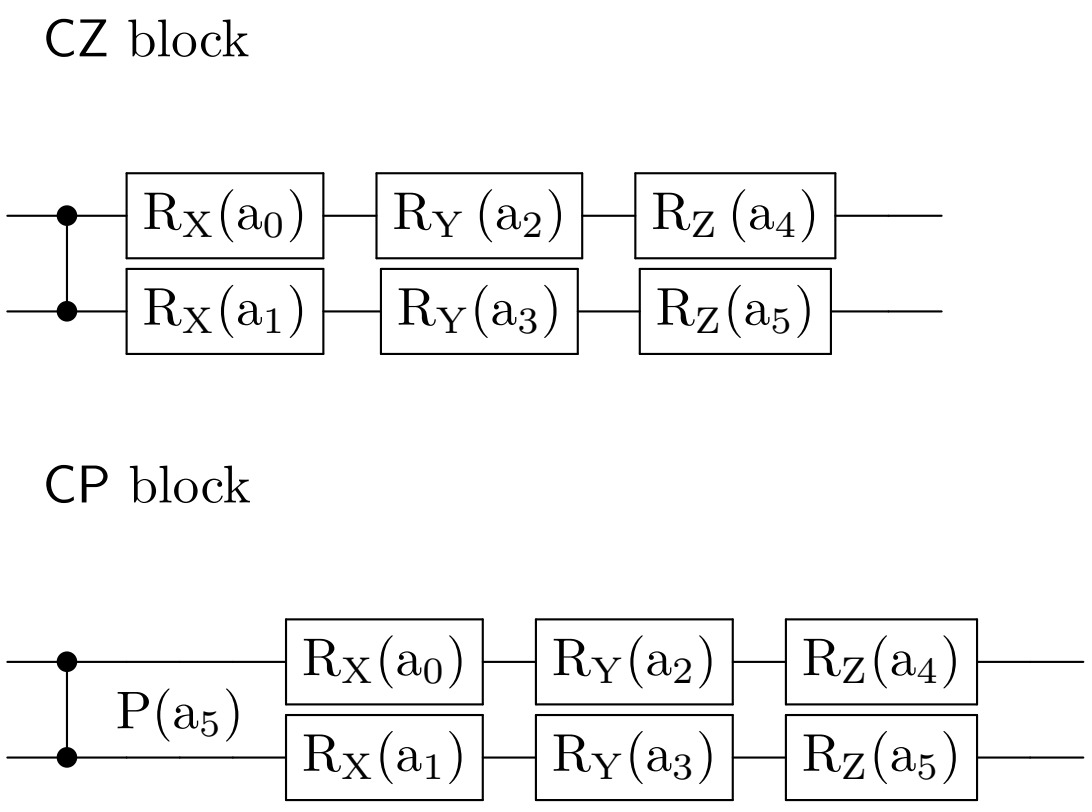

Figure 2: Building blocks for variational templates.

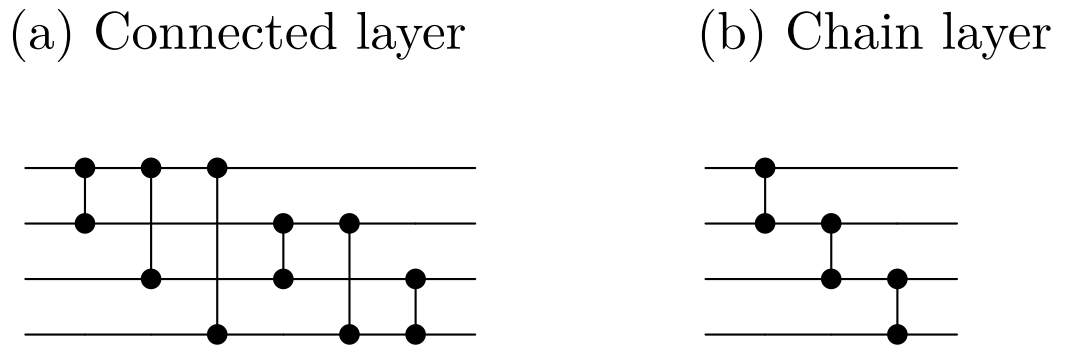

Figure 3: Possible layers for connected and chain (nearest neighbor or line) topologies. Here CZ gates are only meant to schematically specify locations of 2q blocks, not their actual content.

$$D(U,V) = 1 - \frac{|\operatorname{Tr} U^\dagger V|^2}{4^n} \ . \qquad (1)$$

It is normalized to take values between 0 and 1 with the minimum being reached iff $U$ and $V$ differ by a global phase.

## 2.2 Template circuits

The prevalent approach to the variational compiling in the literature [24, 36, 37, 38, 34] that we will follow is to construct the template (or ansatz) circuits by repeated application of two-qubit blocks. The two types of entangling blocks we will use are CZ and CP blocks depicted in Fig. 2. They only differ by the type of the entangling gate used. We will explain the choice of 1q gates shortly. The blocks are further arranged in sequences we refer to as *layers*. In principle, layers can be arbitrary, but we will usually identify layers with coupling maps of the target topology (ordered in an arbitrary way). For example, see Fig. 3 showing layers corresponding to the fully connected and chain (or nearest-neighbor) topology. Finally, to fully specify the template, one must provide the total *number of 2q gates*. Layers are repeated until the specified number of 2q gates is reached, the last layer is truncated if needed. We will write $U^k_{CZ}$ or $U^k_{CP}$ for templates with $k$ entangling gates of type CZ or CP respectively (layer specification is assumed but left implicit in the notation). For illustration, Fig. 4 depicts $U^4_{CZ}$ on a connected topology (here and in the following connected topology means fully connected; connectivity restrictions will always be specified explicitly).

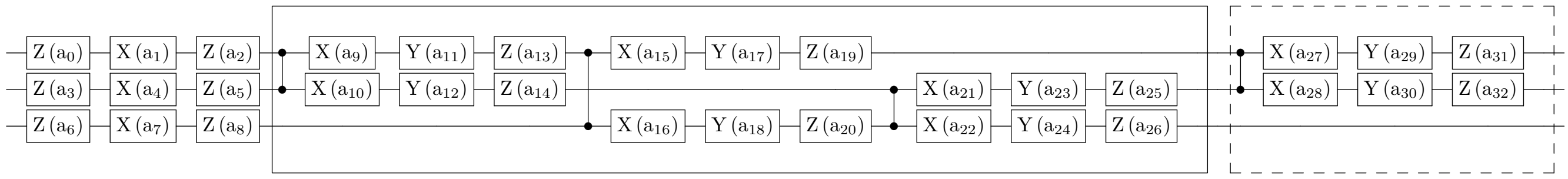

Figure 4: Template circuit $U^4_{CZ}$ on a connected 3q topology with 4 entangling CZ -blocks. The complete connected layer is boxed, the incomplete layer is dash-boxed.

## 2.3 Theoretical lower bound

There is a provable minimum amount of CZ gates required to decompose any $n$-qubit unitary [39], that we will refer to as the *theoretical lower bound*

$$TLB(n) = \frac{1}{4}\left(4^n - 3n - 1\right) \ . \tag{2}$$

It essentially follows from a simple parameter counting argument. We will sketch the argument, which is not only instructive, but also helps to motivate the structure of our template circuits.

A unitary matrix of an $n$-qubit circuit has in general $4^n$ real parameters. Generic 1q gate on the other hand has 3 real parameters, e.g. angles in the Euler decomposition. Thus, without 2q gates, a quantum circuit with $n$ qubits can have no more than $3n$ real parameters. Our template circuits start with a round of 1q gates on each qubit wire, cf Fig. 4. Adding a single CZ gate allows appending two more 1q gates that do not immediately combine with the existing ones. Superficially, this permits adding 6 real parameters per CZ gate. However, one parameter in each 1q gate is redundant, as illustrated in Fig. 5. Using ZXZ decomposition of an arbitrary single-qubit block $U$ and the fact that $R_Z$ commutes with the CZ gate, the leftmost $R_Z$ blocks on each qubit can be pulled to the left and joined with the existing 1q blocks. Therefore, adding a single entangling CZ block allows increasing the real dimension by four. Requiring that the amount of 2q gates is at least sufficient to cover the dimension of the $SU(2^n)$ manifold leads to the equation $4TLB(n) + 3n \geq 4^n - 1$, equivalent to (2) (additional unit subtracted is the irrelevant global phase parameter).

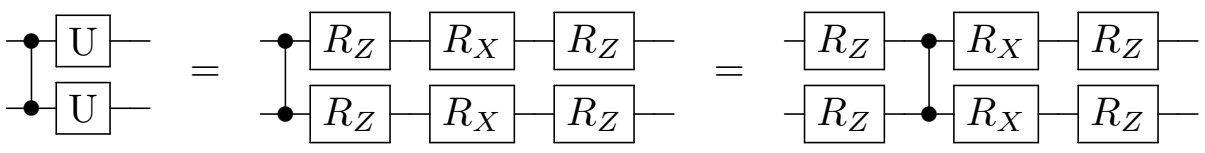

Figure 5: Entangling CZ -block only allows adding four real parameters to the circuit. Explicit gate angles are not depicted.

While expression (2) is a simple theoretical bound, a strong evidence for its tightness exists. First, there is an constructive analytic procedure, known as the quantum Shannon decomposition [40], which synthesizes an arbitrary $n$-qubit unitary using only $\frac{23}{12}TLB(n)$ CNOT gates (roughly twice as much as the theoretical lower bound requires). Second, recent numerical studies [36, 37] suggest that the overhead of the quantum Shannon decomposition is not necessary and that CNOT count given by Eq. (2) is sufficient to compile random unitaries with a great numerical accuracy.

So far, our discussion and the bound (2) addressed generic or random unitaries. However, the unitary matrices of the central importance to quantum computation are highly structured and typically require much less 2q gates. The quantum Shannon decomposition does not appear to be particularly useful in this case. Its extension to restricted topologies is also difficult, often leading to a large multiplicative overhead [40]. Note that being able to find truly optimal decompositions of arbitrary unitaries would amount to determining their gate complexities, which is an NP-complete problem [41]. It is therefore natural to use numeric optimization and heuristic methods in the search for efficient decompositions.

# 3 Variational synthesis and its challenges

As mentioned in the introduction, it is natural to split the variational compiling into the discrete architecture search and the continuous optimization of 1q gates. The difficulty of the architecture search caused by the combinatorial explosion of complexity is manifest. At the same time, the difficulty of continuous optimization also can not be ignored. It is a non-convex problem and thus can not be solved with guaranties. In practice, it may suffer from a range of problems including local minimums, plateaus, and saddle points. Mathematically, the problem of the variational synthesis is very similar to the classical optimization loop in quantum variational algorithms [42], especially their hardware-efficient [43] and adaptive [26] forms. Here, the two key obstacles are the *barren plateaus* and *local minimums*. The barren plateaus [44] manifest as negligible gradients in large areas of the loss landscape and are usually associated with a large number of qubits or parameters. In our experiments with small-scale quantum circuits, we did not find them to be relevant. On the other hand, the problem of local minimums alone is sufficient to render training of the variational algorithms NP-hard [45]. As our numerical experiments suggest, local minimums constitute a real hindrance to the variational compiling.

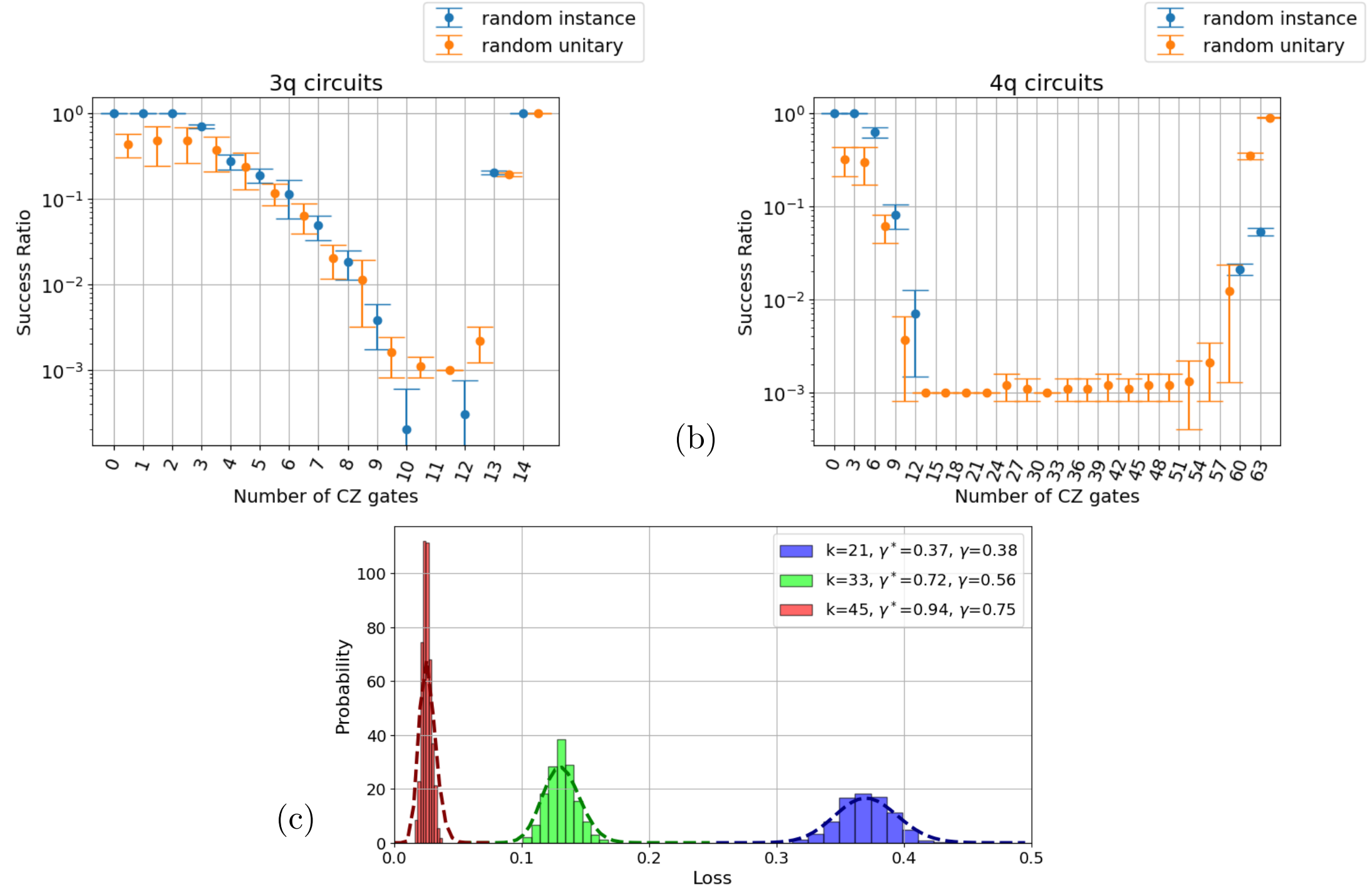

Figure 6: Empirical success ratio as a function of circuit complexity for (a) 3q and (b) 4q templates. Data points for random unitaries are advanced by a half unit along $x$ axis for clarity. (c): Distribution of final loss values for self-instances of 4q templates with depths $k = 21, 33, 45$. Dotted curves are heuristic fits by distribution (4). Expressivity parameters of the empiric fits $\gamma^*$ and computed from the number of parameters in the circuits $\gamma$ are indicated in the legend.

We will quantify the challenges associated with local minimums by the empirical success ratio

$$SR = \frac{M}{N} \,, \tag{3}$$

where $N$ is the total number of times the optimization procedure is performed starting with random initial conditions and $M$ is the number of times the global minimum is reached.

For example, let $U(a)$ be the unitary matrix of the template circuit from Fig. 4 and $a^*$ be some particular choice of angles. It is clear that the global minimum of the Hilbert-Schmidt distance $D(U(a), U(a^*))$ is zero (attained at $a = a*$), but gradient-based optimization does not always reach it (as a cutoff value we take $D \leq 10^{-4}$). With some particular random choice of $a^*$ and random uniform initialization of the template angles, our default optimization (detailed in Sec. 4.3) yields success ratio $SR \approx 0.3$, which implies that roughly two thirds of the times the optimization gets stuck in a local minimum.

We now extend this simple numerical experiment more systematically. Fig. 6 charts the success ratios for 3q and 4q circuits as a function of the number of gates. The basic procedure is the same as above, with several additions. For each gate count $k$, we construct a CZ template with connected layer $U^k_{CZ}$ and find the success ratio of this template learning its random instance $U^k_{CZ}(a) \to U^k_{CZ}(a^*)$. We consider the optimization successful if the Hilbert-Schmidt distance (1) drops below $10^{-4}$. This quite permissive numerical cutoff is enough to reveal the local minimums without worrying about the optimization details such as convergence rate.[1] Independent evidence that unsuccessful attempts are indeed due to local minimums will be presented at the end of the section.

More precisely, we take 10 different template in-stances for each gate count $U^k_{CZ}(a^*_{1-10})$ and compute the success ratio for each of them using 1000 initial conditions, generated uniformly at random. Blue markers represent mean success ratios averaged over 10 target circuits, while error bars quantify the standard deviations. Absence of blue markers implies that the empirical success ratio turned out vanishing, i.e. that the global minimum was not reached. For 4q circuits data points were only collected for $k = 3n, n \in \mathbb{N}, k \leq 63$.

[1]In later sections, when compiling unitaries of interest, we will typically impose a stricter cutoff $D < 10^{-6}$, which approaches the machine precision of our setup.

There are several remarkable features of these plots. First, the success ratio drops very quickly as the 2q gate count increases, reaching values below $10^{-3}$ at 10 CZ gates for 3q circuits and 15 CZ gates for 4q circuits. Next, perhaps surprisingly, the success ratio rises back to values of order 1 as the number of CZ gates approaches the theoretical lower bound (2). In fact, this is in agreement with the empirical evidence found in the literature [36, 37, 35] that near the theoretical lower bound numerical compilation appears to be very efficient as if the problem was convex. That over-parameterized quantum circuits can often be trained efficiently have also been motivated theoretically [46, 47]. Finally, although there is a certain spread of success ratios across different template instances, dependence on the 2q gate count sets the dominating trend.

This suggests that the success ratio is mostly determined by the template, not by the target. To confirm this intuition, we carried out additional experiments using random unitaries $V$ instead of template instances as targets. The difficulty here is that the true value of the global minimum of $D(U^k_{CZ}(a), V)$ is not known, yet the presence of local minimums is still manifest, because different optimization runs tend to end up with significantly different loss values. We modify the definition of the success ratio in this case, by counting as successful all optimization runs that approached sufficiently closely the lowest value across all runs for a given target unitary (using the same cutoff as before $D - D_{min} \leq 10^{-4}$). Note that with this modified definition, the success ratio can never be zero (because there is always at least a single run with the lowest value). We see that in the regime when success ratios for random instances are sufficiently high, success ratios for random unitaries closely parallel them, both in mean and in deviation. In the regions where success ratios for random instances are very small or vanishing, success ratios for random unitaries are nonzero (they can not be by construction) but are close to zero. We expect them to drop further if more samples are accounted for. Overall, our experiments strongly suggest that local minimums are mostly determined by the templates and not by the targets.

It is also instructive to inspect not just the success ratios, but the distribution of loss values for different templates. The histograms at Fig.6(c) depict distributions of final loss values achieved by the optimization starting from random initial conditions for connected 4q templates $U^k_{\mathsf{CZ}}(a)$ with three different depths $k$. The first observation here is that most loss values are clustered near a mean value away from the global minimum (by construction, the global minimum has zero loss). Next, the quality of the local minimums increases as the depth (and hence expressivity) of the template grows, and the spread shrinks.

This is in a remarkable agreement with recent analytic results [48, 49], where the following asymptotic distribution of the density of critical points $E_0$ for Hamiltonian-agnostic variational loss functions was derived

$$E_0 \sim e^{-mE/2} E^{l/2-m} (1-E)^l \ . \tag{4}$$

Here $E$ is the variational loss function normalized to satisfy $0 \leq E \leq 1$, $m$ the dimension of the Hilbert space, and $l$ the number of independent parameters. Ratio

$$\gamma = \frac{l}{2m} \tag{5}$$

quantifies the expressivity of the circuit and crucially affects the distribution of critical points. For $\gamma \ll 1$ most local minimums are far away from the global minimum and the loss function is hard to train. For $\gamma > 1$ the local minimums cluster exponentially close to the global minimum and the model is easy to train (yet it is of exponential depth).

In Fig.6(c) we fit the loss histograms with the distribution (4). Using the Hilbert-Schmidt test (see e.g. [24]), the compilation problem on $n$ qubits can be reformulated as the state preparation problem on $2n$ qubits. Hence we choose the dimension of the Hilbert space $m = 2^8$. The expressivity parameters $\gamma^*$ are fitted in an ad hoc way, to visually match the histograms. The expressivity parameters $\gamma$ (5) are also indicated in the plot, but distributions corresponding to $\gamma$ do not align well with the histograms and are not shown. Of course, one should not expect a precise quantitative agreement between the distribution (4) and our empiric histograms. For one, (4) is an asymptotic statement valid for large system sizes. There are other assumptions going in the derivation of (4) that our setup may fail to satisfy. Nevertheless, we view the qualitative agreement between our empiric results and the theoretical analysis as a strong indication that the local minimums in the variational compilation are real, present a significant hindrance, and should be taken into account in any synthesis approach relying on numerical optimization of parametric gates.

Interestingly, Figs. 6(a,b) suggest that for the circuits with few parameters, the local minimums are not present. Also, we note that for the 4q circuits, success ratios initially drop more slowly than for 3q circuits. Studying the onset of local minimums, and it's scaling with the system size, is an interesting question that we do not address here.

We should mention that a success ratio is not only a function of the loss landscape, but also of the optimization algorithm and the distribution of the initial conditions. One popular optimization mode for quantum algorithms is a layerwise training [50], as it uses less computational resources and can mitigate the Barren plateaus [51]. However, these concerns are not relevant on the scale of our experiments and, moreover, the layerwise training is known to raise additional trainability issues in some cases [52]. There is a number of proposals to alleviate the problem of local minimums by the choice of optimizer [53, 54], but in our experiments, none performed sufficiently better than simple ADAM [55]-based optimization to justify additional computational resources that are typically required by higher-order methods such as the natural gradient [56] or imaginary time evolution [57]. A recent empirical comparison of various optimization methods for quantum variational algorithms [58] also suggests that ADAM optimizer is often the simplest and most efficient choice.

In contrast, our experiments suggest that parametrization and/or distribution of the initial parameters can have noticeable effects. As explained in Sec. 2.3 template structure illustrated at Fig. 4 features redundant 1q gates. In any entangling block, two rotation gates can be removed without compromising circuit expressivity, i.e. without any shrinking in the space of all unitaries obtainable from the template. However, performance of the templates with the minimal number of 1q gates appears to be worse on average (though there are counter examples, see Sec. 5.2). This can be due to the fact that overparametrization favorably deforms the loss landscape and/or because the random uniform initialization of the angles produces a different distribution of the initial unitaries. In the present study, we do not attempt to disentangle the two possible effects and leave this important question for future work. Unless stated otherwise, reported results correspond to the 'XYZ' templates as per Fig. 4.

# 4 The CPFlow algorithm

## 4.1 Motivation and overview

In the context of variational synthesis, results of the previous section suggest that solving the continuous optimization problem may be just as difficult as solving the discrete architecture search: even if the structure of the template is a perfect match for the target unitary, finding the suitable angles may be very challenging. In the absence of an efficient way to solve the latter problem in our approach, we choose the brute force route of an extensive multi-start optimization.

Our second main technique is to relax the discrete architecture search to yet another continuous optimization. For illustration, consider the circuit at Fig. 7. Here, the 2q gates are the controlled phase gates (1), which interpolate between the identity gate $\mathsf{CP}(0) = \mathbb{I}$ and the CZ gate $\mathsf{CP}(\pi) = \mathsf{CZ}$. For generic values of the angle, a single $\mathsf{CP}(a)$ gate can be decomposed into 2 CZ gates (plus 1q gates). Therefore, different values of parameters in CP gates in the template (7) effectively capture several different templates with the 2q CZ gates and training templates with the CP blocks can encompass both the architecture search and the tuning of continuous parameters, moreover performed in a coherent manner.

We can anticipate, however, that training CP templates directly will result in most CP gates having generic angles and hence effectively doubling the CZ count of the original template. To address this issue, we introduce an additional *penalty term* to the loss function that is intended to drive all CP angles to either 0 or $\pi$. The shape of the penalty function that we use is presented in Fig. 8.

This penalty is intended to drive all CP angles during the optimization to either 0 or $\pi$, and hence to reduce the CZ count of the resulting circuit. Moreover, at values $a = 0, \frac{1}{2}\pi, \frac{3}{2}\pi, 2\pi$ the regularization term faithfully captures the CZ cost of the CP gate. We choose a simple linear interpolation between these values because piecewise-linear penalty functions are known to lead to the discrete decision-making in certain cases of continuous relaxation of discrete optimization problems such as sparcification of machine learning models [59], compressed sensing [60, 61], and robust PCA [62] to name a few. For numerical stability, small plateaus near the reference values are added (empirically we find that CP angles often relax near $\frac{1}{2}\pi, \frac{3}{2}\pi$ as well).

An obvious problem with this regularization function is the presence of the local minimum at $a = \pi$. In fact, enumerating all local minimums associated with the regularization terms of a CP template is equivalent to the discrete search through all CZ templates that it can reduce to. However, our empirical results suggest that simultaneous optimization over CP and 1q angles is a very efficient strategy if the overall weight of the regularization term is chosen properly. If it is too small, the regularization term has little effect and the resulting decompositions tend to have a high CZ count. When the weight is too high, the CP angles effectively get captured by the closest local minimum and the flexibility of our strategy is lost. In fact, optimization with a high regularization weight could be considered a version of a random search over the architectures (if the initial CP angles are chosen randomly) and performs significantly worse than optimization with a properly tuned regularization weight.

## 4.2 Procedure

We formulate the general synthesis problem as follows. Let $L(U)$ be the loss function to be minimized with unitary as the argument. For unitary synthesis $L(U)$ may be any measure of fidelity to the target unitary $V$, for example $L(U) = D(U, V)$. For state preparation, one can choose $L(U) = |\langle\psi|U|0\rangle|^2$ where $|\psi\rangle$ is the target state and $|0\rangle$ is the usual reference state. Other loss functions can be used as well, for an example of compiling the unitary up to a diagonal multiplier see App. B. The goal is to find a unitary $U^k_{CZ}(a)$ such that $L(U^k_{CZ}(a))$ is sufficiently close to the global minimum of $L(U)$ and at the same time the number $k$ of 2q gates is as small as possible.

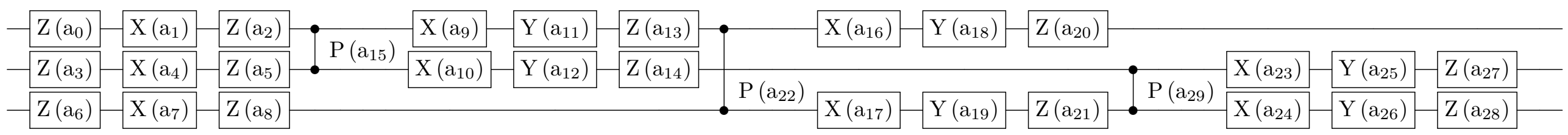

Figure 7: Template 3q circuit $U^3_{CP}$ on a connected topology.

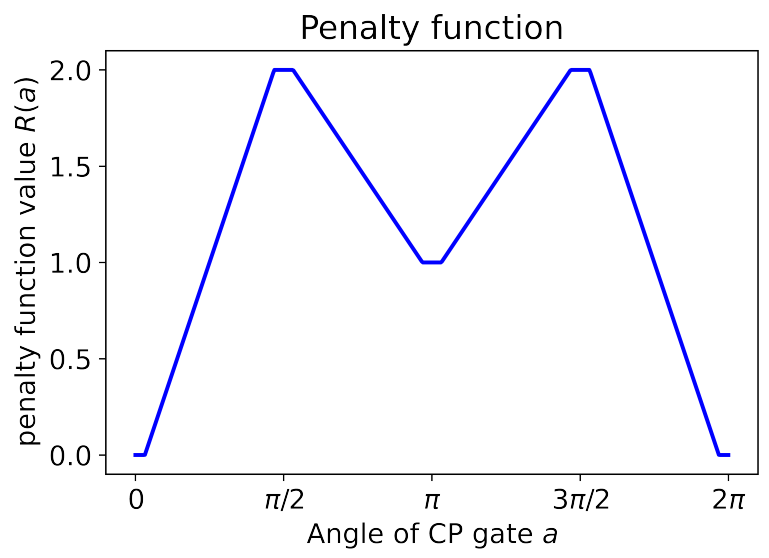

Figure 8: Penalty function for angles of the CP gates. For clarity of the figure, the width of plateaus near $0, \frac{1}{2}\pi, \frac{3}{2}\pi, 2\pi$ is exaggerated.

The regularized loss optimized by CPFlow reads

$$\mathcal{L}(a) = L(U^k_{CP}(a)) + r \sum_{a_i \in CP} R(a_i)\ . \qquad (6)$$

The first term is the value of the original loss function evaluated with the variational circuit as the argument. The second term is the regularization term, summing the CP penalties for all CP angles in the template. The number of 2q gates $k$ and the overall regularization weight $r$ are two of the most important hyperparameters of the model.

Three main stages of the algorithm are described below. Precise details are given in Sec.4.3.

## Static synthesis

1. ***Raw sampling.*** Loss function (6) is minimized starting from many initial conditions (`num_samples`). For each sample, both the CP angles and the angles of 1q gates are generated uniformly and independently at random.

2. ***Selecting prospective results.*** Results of the first step are filtered based on two criteria (i) the original loss function $L(U^k_{CP})$ must be below a given threshold (`entry_loss`) and (ii) the number of CZ gates in a projected CP circuit must be below a specified value (`accepted_num_cz_gates`). Condition (i) means we only accept circuits that are close enough to the global minimum, while (ii) rejects decompositions with too high CZ count. Projection from CP to CZ circuits $U^k_{CP}(a) \to U^{k'}_{CZ}(a')$ is preformed by rounding off angles of CP gates that are sufficiently close (within `threshold_cp`) to 0 or $\pi$ and substituting other CP gates with their CZ decompositions.

3. ***Verification.*** At this stage the projected circuits contain only CZ gates and the regularization term is removed. For each prospective CZ circuit, the original loss function $L(U^{k'}_{CZ}(a))$ is further optimized starting from initial angles $a'$ inherited from the CP circuit. The verification is considered successful if the CZ circuit reaches a more stringent loss threshold (`target_loss`).

This basic scheme can be modified in many ways: by choosing a different regularization function, different sampling of the initial angles or altering the details of the gradient based optimizer to name a few. We have mostly experimented with varying two hyperparameters that are evidently crucial, the number of CP gates $k$ and the regularization weight $r$. Heuristically, we find that a reasonable number of CP gates is usually between $k_0$ and $2k_0$, where $k_0$ is the expected optimal CZ count of the decomposition. A performant choice for the regularization weight $r$ for loss functions normalized so that $0 \leq L(U) \leq 1$ is $r = 5 \times 10^{-4}$. There could be exceptions to both these rules of thumb. To make better choices of hyperparameters on a case by case basis, we use the Bayesian tuning algorithm provided by the Hyperopt package [63].

Tuning of hyperparameters is significantly hindered by the fact that the loss function is stochastic. Taking sufficiently many samples to reliably estimate the quality of a hyperparameter configuration may cost too much computational resources, while not taking enough can make the acquired data too noisy to be useful. On the positive side, the ultimate goal is not to find the best hyperparameters, but rather to find the best decompositions which routinely occur at suboptimal points as well. The routine including hyperparameter tuning can be summarized as follows.

**Adaptive synthesis**

1. ***Defining the search space.*** Choose distributions to draw the number of gates $k$ and the regularization weight $r$ from. Typically, we use uniform distribution for $k$ in some integer range and lognormal distribution for $r$ with mean around $5 \times 10^{-4}$ and standard deviation 0.5.

2. ***Evaluating the score function.*** Draw a sample $k, r$ from the hyperparameter distribution according to the Hyperopt algorithm. Execute steps 1 and 2 from the STATIC routine. Any CZ count is accepted at this stage, i.e. the results are only selected by the value of the original loss function $L(U^k_{CP}(a))$. Let $k_1, k_2, \dots$ be CZ counts of all prospective results selected. We define the score function (reminiscent of *softmin*) by

$$\text{score} = -\log_2 \left( \frac{1}{N} \sum_i 2^{-k_i} \right) , \qquad (7)$$

where $N$ is the total number of raw samples. The intuition is as follows. Templates that are too expressive and have high regularization weight will yield many high-fidelity decompositions with excessive CZ counts. Imposing stricter hyperparameters will yield more efficient decompositions, but also more raw samples failed to converge.

   Function (7) balances between these scenarios by averaging CZ counts of accepted decompositions, weighting them exponentially, i.e. a single decomposition with $k$ CZ gates scores as two decompositions with $k+1$ or four with $k+2$. The minimum of this function is $k_0$, the CZ count of the best possible decomposition. It could be achieved if all raw samples reach the threshold fidelity and have the optimal CZ count $k = k_0$. If some of the raw samples fail to converge or require higher than optimal CZ count, the score function increases. The maximum score $= +\infty$ is obtained when none of the raw samples passed the fidelity threshold.

3. ***Verifying best decompositions.*** At the previous stage, prospective decompositions are usually not verified as the verification process is time-consuming and should not significantly alter the score estimation (some of the decompositions may fail to pass the verification, but this is rare). However, if there are prospective decompositions that improve on the current best they are verified and if accepted, the current best is replaced.

4. ***Repeat.*** Repeat steps 2 and 3 until either the maximum number of score evaluations is reached or a decomposition with the desired number of gates is found.

As the result of the ADAPTIVE routine, one narrows down the space of good hyperparameters for the problem and collects several efficient decompositions found in the process. This may already be sufficient for the end goal, or provide a good starting point to generate more decompositions using the STATIC routine with appropriately chosen hyperparameters. Although the algorithm directly targets only minimization of the CZ gate count, generating many similar decompositions allows one to further select by other criteria such as CZ depth, or even T count and T depth. We will illustrate this process in Sec.5.1.

## 4.3 Technical details

The STATIC routine implemented in CPFlow proceeds as follows. First `num_samples` of initial angle combinations are generated uniformly at random. Learning at the raw sampling stage proceeds with the `ADAM` optimizer with `learning_rate` 0.1 ran for `num_gd_iterations` (2000 by default). At the selection stage for each sample, the best configuration of angles $a^*$ is chosen corresponding to the minimum of the regularized loss function (6) across all iterations. If the primary loss function at this configuration of angles $L(U^k_{CP}(a^*))$ is below the `entry_loss` threshold ($10^{-3}$ by default) the CP circuit is projected to the CZ circuit $U^k_{CP}(a^*) \to U^{k'}_{CZ}(a')$. Projection is performed as follows. CP gates with angles lying within the `cp_threshold` (0.2 by default) distance away from 0 or $\pi$ are replaced by the identity and CZ gates respectively. CP gates that were not replaced at the previous stage are decomposed into 2 CZ decomposition using standard methods. If the 2q gate count $k'$ of the resulting CZ circuit is below `accepted_num_cz_gates`, the circuit is deemed prospective.

At the verification stage, the original loss function $L(U^{k'}_{CZ}(a))$ is further optimized with the `ADAM` optimizer with a smaller `learning_rate_at_verification` (0.01 by default) ran for an increased number of iterations `num_gd_iterations_at_verification` (5000 by default). Importantly, the new optimization starts with the initial angles $a'$ obtained after projecting the CP circuit. If the new optimization pass reaches a more stringent `target_loss` threshold ($10^{-6}$ by default) the verification is considered successful and the resulting circuit is added to the collection of decompositions.

The parameters specified above work well with the Hilbert-Schmidt loss function (1) in a sense that the majority of prospective decompositions pass at the verification stage. For other normalizations/shapes of the loss function, different parameter specifications might be required.

The ADAPTIVE search basically consists of several STATIC rounds. At each round the score function (7)

is computed from the prospective results. If there is a prospective result that improves the current best decomposition, it is verified and added to the decomposition pool if successful. Otherwise, the verification stage is omitted and a new STATIC round with altered hyperparameters is initiated. The total number of rounds is controlled by `max_evals` (100 by default). Hyperparameters for each round are chosen by the Hyperopt implementation of the tree-structured Parzen Estimator algorithm [63] based on the previous score evaluations and input parameter distributions specified by the user. We use the uniform distribution for the number of gates $k$ in the range between `min_num_cp_gates` and `max_num_cp_gates`. For the regularization weight $r$ we use lognormal distribution with default values `r_mean`$=5.5\times10^{-4}$ and `r_variance`$=0.5$. Note that by the default that we keep, the first 20 parameter choices in Hyperopt are intended to sample broadly from the search space and do not depend on the previous score evaluations (only on the input distributions), the actual optimization starts at further steps. However, with a good choice of the initial distributions optimal or near optimal decompositions are often found by CPFlow within several first evaluations.

### 4.4 Computational setup

CPFlow [64] is written entirely in Python, with the computational efficiency enabled by the JAX library [65]. One advantage of CPFlow is that the computations are highly parallelizable as the core of both basic routines consists in performing independent multi-start optimizations. We also rely on Qiskit [7] for visualization, validation, and post-processing. Numerical experiments reported in this paper were carried out on a server equipped with a 16 GB NVIDIA Quadro RTX 5000 GPU. A single STATIC routine with 1000 samples for 4q and 5q unitaries took several minutes in our setup. Correspondingly, a typical ADAPTIVE routine with 100 evaluations using 1000 samples each took several hours.

## 5 Synthesis of Toffoli gates

We now put the CPFlow algorithm to work. We choose the Toffoli gates as our key benchmark examples, motivated by several considerations. First, the Toffoli gates are among the most essential building blocks for a great variety of quantum algorithms. Next, large multi-controlled Toffoli gates can be built recursively from the smaller ones [9], hence optimization of the small-size Toffoli gates can potentially be propagated to the large multi-controlled gates required in useful quantum algorithms. Lastly, Toffoli gates have been studied extensively [9, 66, 67, 68, 69] and although deriving rigorous bounds beyond the 3q case is very difficult, the best existing decompositions

| | Best number of CP gates $k$ | Best regularization weight $r$ | Optimal decompositions |
|---|---|---|---|
| Connected | 7 | $1.31\times10^{-3}$ | 28/100 |
| Chain | 14 | $0.88\times10^{-3}$ | 19/100 |

Table 1: Synthesis statistics for the 3q Toffoli gate.

are likely to indeed be optimal and hence provide a perfect benchmark.

The basic Toffoli gate, also known as the Controlled-Controlled-NOT or the $C^2X$ gate, is depicted as follows.

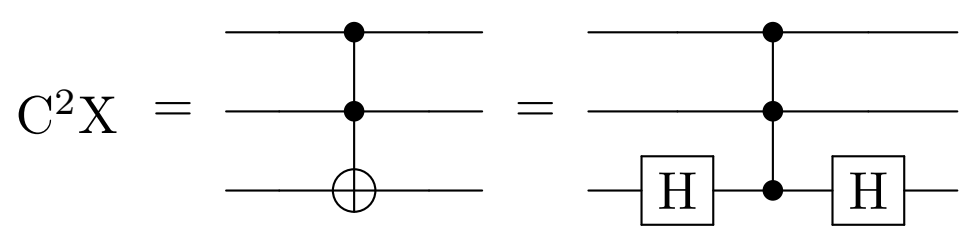

The right diagram represents the Toffoli gate as the $C^2Z$ gate conjugated by the two Hadamard gates ($HZH = X$). The $C^2Z$ gate itself is represented by a diagonal matrix $C^2Z = \mathrm{diag}(1,1,1,1,1,1,1,-1)$ and is symmetric with respect to all qubits. Therefore, up to a conjugation by 1q gates the Toffoli gates ($C^2X$ as well as any $C^nX$) are also symmetric. This implies that even if the symmetry between the qubits is broken by e.g. a non-trivial topology, the choice of the target qubit for the Toffoli gate is not relevant for our synthesis problem.

### 5.1 3q Toffoli

In this subsection, we use CPFlow to find efficient decompositions of the 3q Toffoli gate and illustrate many important features of the algorithm along the way. 3q Toffoli gate can be decomposed into 6 CZ gates on the fully connected topology or 8 CZ gates on the chain topology. Using CPFlow we were able to find many inequivalent decompositions with these optimal CZ counts. First, we ran the ADAPTIVE routine with 100 evaluation steps to identify the best hyperparameters in each case. Results are reported at Fig. 9. Best hyperparameters for each topology are shown in Table 1.

Note that at this stage decompositions with the optimal CZ counts have already been found, but it is instructive to further analyze the performance of the algorithm and generate more decompositions. To this end, we ran the STATIC routine with optimal hyperparameters and 100 samples for each topology. The third column in Table 1 states how many of the initial conditions led to the decompositions with the optimal CZ count. For connected and chain topologies, chances of finding optimal decompositions are roughly 30% and 20% respectively. These figures are to be compared with the success ratios at the corresponding gate counts shown at Fig. 6, which are of order 10% and 2% respectively. The remarkable conclusion is that (at least in this simple example) our strategy of coherent learning of the architecture together with continuous parameters seems to be unreasonably efficient. With the best hyperparameters, the frequency of finding CZ count optimal decompositions is greater than the mean success ratios for reaching the global minimum for fixed architectures.

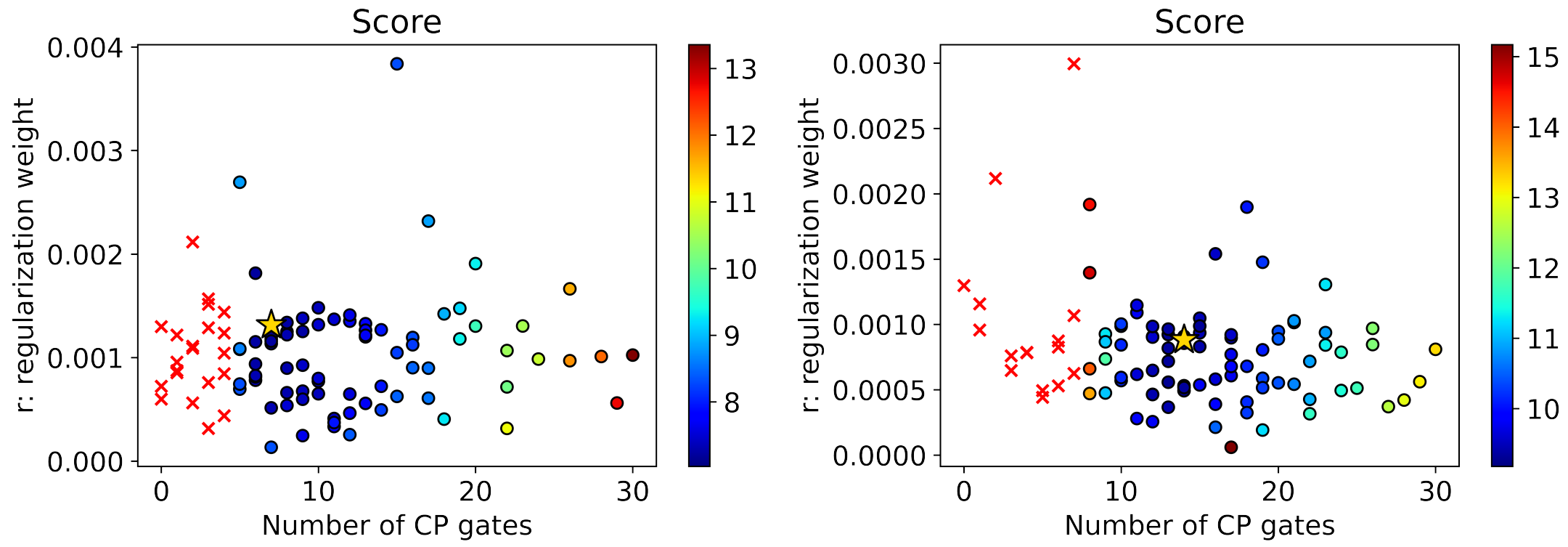

Figure 9: Visualization of the hyperparameter optimization during ADAPTIVE synthesis of the 3q Toffoli gate on connected (left panel) and chain (right panel) topologies. Red crosses corresponds to infinite score values and imply that no valid decompositions were found at these points. Gold stars mark the best hyperparameter configurations.

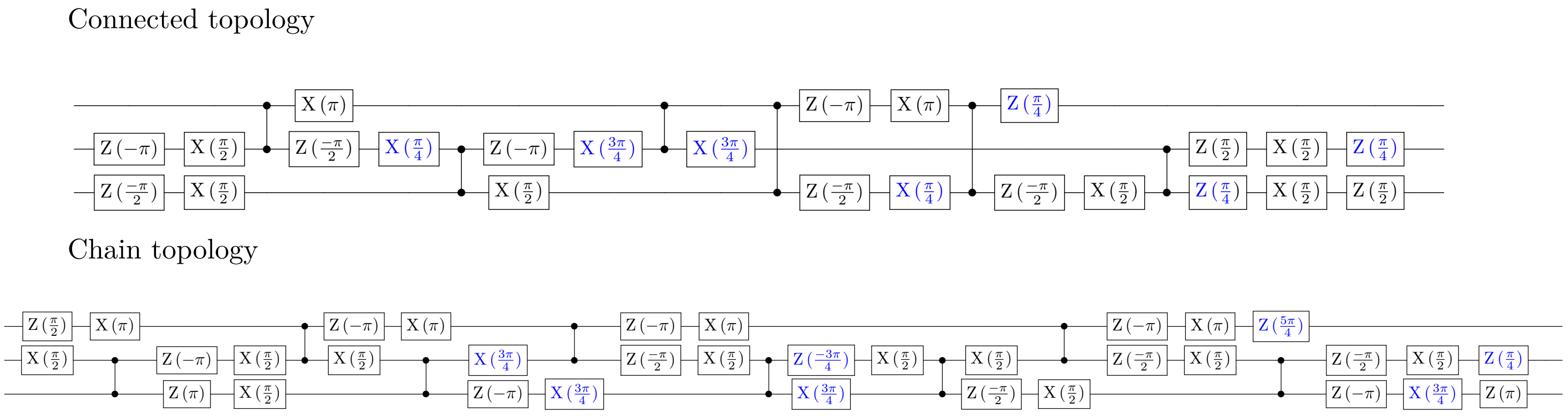

Figure 10: Decompositions of the 3q Toffoli gate that are likely to be optimal with respect to all four metrics: CZ depth, CZ count, T depth, T count. Rotation gates are shortened from $R_\sigma$ to $\sigma$ for readability. Non-Clifford gates, each obtainable from a single T gate, are highlighted.

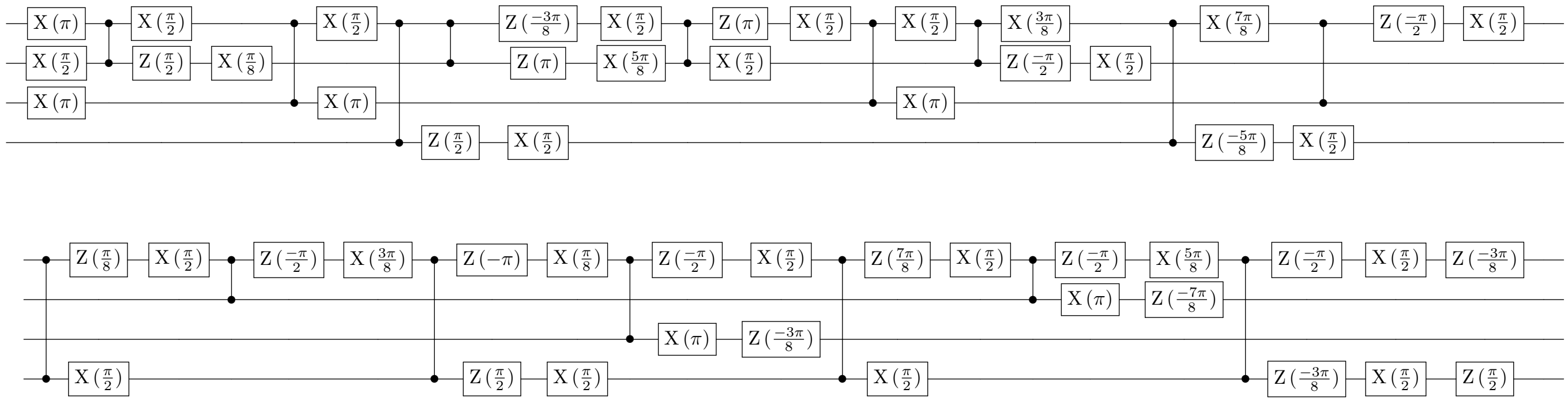

Figure 11: Decomposition of the 4q Toffoli gate on the star-shaped topology (all CZ gates touch the uppermost qubit) with 16 CZ gates.

Decompositions which are initially generated by CPFlow are approximate and bear little resemblance to the standard Clifford+T representations, where 1q gates are expressible as rotation gates with angles being rational multiples of $\pi$. We develop a simple procedure that attempts to eliminate redundant gates, rationalize remaining ones, and if possible expand the circuit into Clifford+T gate set. Details of the procedure are deferred to App. A. The procedure is heuristic and does not guarantee elimination of all spurious angles, but often works well in practice. Applying it to the decompositions found at the STATIC stage, we were able to generate 12 Clifford+T decompositions of 3q Toffoli gates for each topology. We then sorted these decompositions according to four metrics : CZ count, CZ depth, T count and T depth. For both topologies decompositions with the smallest T depth were simultaneously optimal with respect to the three remaining metrics. Fig. 10 depicts the best decompositions that were generated. Note that the decomposition on the chain topology has T depth 3 and is possibly a new result.

By construction, the distance $D$ of decompositions reported in Tab. 1 and depicted in Fig. 10 to the exact 3q Toffoli gates satisfies $D < 10^{-6}$, approaching the machine precision in our setup. It is natural to assume, that decompositions with rational angles, like those in Fig. 10, are in fact exact. Since all matrix elements of the rationalized decompositions are polynomials in $e^{i\pi p/q}$ $(p, q \in \mathbb{Z})$ with integer coefficients, it must be possible to reduce the unitary of the decomposition to the target unitary using basic algebraic manipulations. Using symbolic algebra software external to CPFlow we verified that circuits in Fig. 10, as well all other decompositions presented explicitly in the paper (Figs. 11,15,14) are exact. Directly integrating this verification into CPFlow is left for future work.

## 5.2 4q Toffoli

We now proceed to the decomposition of the 4q Toffoli gates, which are significantly more challenging. Variational synthesis of the 4q Toffoli gate on various 4q topologies has been recently addressed in Ref. [38]. The approach adopted there was that of an exhaustive search over all architectures. It is interesting to note that for exhaustive search connectivity restrictions actually simplify the problem. Using CPFlow we were able to reproduce all results presented in Ref. [38], see Table 2. Moreover, for the star-shaped topology we achieved a minor improvement, reducing the CZ count from 17 to 16. The corresponding circuit is depicted at Fig. 11. Note that we did not look for decompositions minimizing CZ depth, but simply report CZ depths of the first CZ count optimal decompositions found during the search.

| Topology | 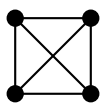 | 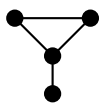 | 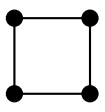 | 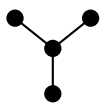 | 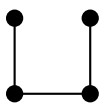 |
|---|---|---|---|---|---|
| CZ count | 14 | 14 | 16 | 16 | 18 |
| CZ depth | 11 | 13 | 15 | 16 | 14 |

Table 2: Decompositions of 4q Toffoli gate on various topologies

| | XYZ | XZ |
|---|---|---|
| 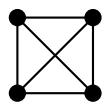 | $0.6 \times 10^{-2}$ | $7.8 \times 10^{-2}$ |
| 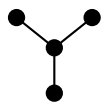 | $0.4 \times 10^{-2}$ | $0.2 \times 10^{-2}$ |

Table 3: Empirical success ratios for optimal decompositions of the 4q Toffoli gate determined from 500 samples.

In all except for the fully connected topology, decompositions were discovered by the ADAPTIVE algorithm with the full range of template depth for 4q unitaries (0, 61), 500 samples at each hyperparameter configuration, and 50 hyperparameter evaluations. The clock time taken by the search for each topology was about 40 minutes on our server 4.4. In contrast, the optimal decomposition on the fully connected topology only appeared after about 200 hyperparameter evaluations and took about 2 hours.

One might wonder why did the exhaustive search approach of Ref. [38] overlook the decomposition on the star topology with 16 CZ gates. A plausible cause might again be due to the local minimums. Table 3 shows empirical success ratios for the optimal circuits found by CPFlow on the fully connected and star topologies with two different choices of 1q gates: XYZ and XZ. First thing to note is that the success ratio for the star topology and XZ structure of 1q gates is only about 2%. Work [38] indeed used the XZ template, albeit with a different optimization procedure [70]. We find it likely that the appropriate architecture was missed simply due to an insufficient amount of trials. In fact, the problem of local minimums deprives of guarantees even the exhaustive search over architectures Another interesting observation is that for a connected circuit the success ratios of XYZ and XZ templates differ by an order of magnitude. This highlights the importance of the parametrization and/or initial sampling.

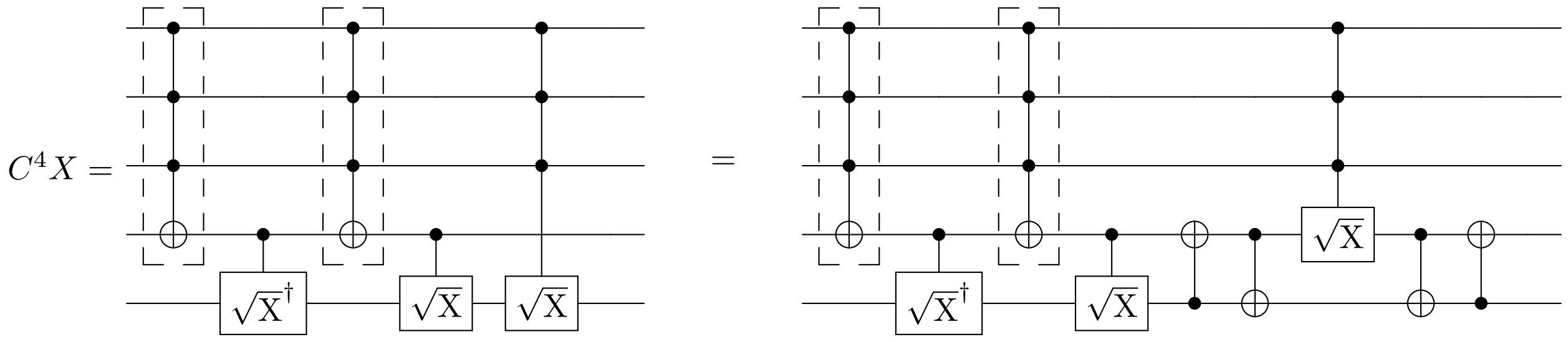

Figure 12: A decomposition of the 5q Toffoli gate.

### 5.3 5q Toffoli

To our knowledge, the best decomposition of the 5q Toffoli gate on the fully connected topology without ancilla qubits features 30 CZ gates, an upper bound valid for all diagonal gates [40]. The best result of direct synthesis with CPFlow running for several hours that we observed featured 36 CZ gates, indicating that 5q unitaries with sufficiently many gates are significantly harder to address. However, synthesis with topological restrictions also poses significant challenges to the demultiplexing framework of Ref. [40]. It is thus interesting to see if our numerical routine can be useful.

To this end, we consider the compilation of the 5q Toffoli gate on the chain topology. As a reference point we take the best result achieved by the Qiskit transpiler out of 1000 runs with `optimization_level` set to 3, that yielded a decomposition with 61 CZ gates.[2] The best results of direct synthesis with CPFlow yielded a decomposition with 69 CZ gates. As we now show, by using a combination of the analytic and numerical techniques this result can be improved to 48 CZ gates.

Our strategy is to reduce synthesis of the 5q Toffoli gate to the synthesis of several 4q blocks. We first observe that the optimal 30 CZ gate decomposition of the 5q Toffoli gate on a fully connected topology can be obtained from the left circuit depicted in Fig. 12. Triply controlled $\sqrt{\mathrm{X}}$ gate is a diagonal gate up to a conjugation by the Hadamard gates and hence can be decomposed using 14 CZ gates, just as the standard 4q Toffoli gate. Singly controlled $\sqrt{\mathrm{X}}$ gates can be decomposed into 2 CZ gates each. Finally, the boxed 4q Toffoli gates can be replaced by their relative phase counterparts [67], each requiring only 6 CZ gates. In total, this gives 30 CZ gates, the best known amount.

We now try to adapt this decomposition to the chain topology. The right circuit in Fig. 12 shows that by inserting four additional CNOT gates (each pair originates from a SWAP gate, the two closest CNOT gates cancel each other) around the triply controlled $\sqrt{\mathrm{X}}$ gate we can place all 4q gates on the first four qubits. Next, we use CPFlow to decompose $\mathrm{C}^3\sqrt{\mathrm{X}}$ and relative phase $\mathrm{C}^3\mathrm{X}$ gates on the chain topology. We found a decomposition of $\mathrm{C}^3\sqrt{\mathrm{X}}$ with 18 CZ gates, the same gate count that is needed for decomposing of the $\mathrm{C}^3\mathrm{X}$ gate on a chain topology 2. For the relative phase $\mathrm{C}^3\mathrm{X}$ gate, we found a decomposition with 11 CZ gates. Details and corresponding circuit diagrams are delegated to App.B. In the end, this yields a decomposition of the 5q Toffoli gate on the chain topology with $48 = 2 \times 11 + 18 + 2 \times 4$ CZ gates in total. We are not aware of other methods improving this count.

## 6 Further benchmarks

Following [34], we test the performance of CPFlow on a range of standard benchmark circuits from the ibm_qx database [72, 71]. In Ref. [34] an extensive comparison between packages SQUANDER , QFast , and QSearch was performed. Provided enough time, SQUANDER reliably outperformed other packages in most examples. From each of the Tables 1, 3 and 4 presented in Ref. [34] we pick 5 circuits with the highest CZ count found by SQUANDER , as these are likely to be the most challenging and hence the most informative (for the same reason we skipped Table 2, which mostly contains much simpler circuits). All selected examples are 5q circuits. As before, we consider the compilation successful if the distance to the target unitary (1) is less than $10^{-6}$, which is approaching the machine precision in our setup.

We must note that currently CPFlow does not have a dedicated subroutine to estimate the expected target complexity and narrow the hyperparameter window accordingly. Using the ADAPTIVE routine with the full range of gate counts allowed by the theoretical lower bound (2) already for 5q circuits is unnecessarily time-consuming. Informed by the gate counts obtained by SQUANDER we ran the ADAPTIVE routine with CP counts of templates in the range from

[2] Note that the transpilation process in Qiskit generally yields circuits which are only equal to the target unitary up to a possible permutation of qubits. As restoring the original permutation might be an expensive operation in terms of the CZ count, the transpilation results should be considered as a lower bound.

| Circuit | CPFlow | SQUANDER | QSearch | QFast+Qiskit+SQUANDER |
|---|---|---|---|---|
| I. Connected, lower complexity | | | | |
| 4gt5_76 | 21 | 24 | - | - |
| one-two-three-v2_100 | 28 | 37 | 43 | - |
| alu-v3_34 | 14 | 25 | 27 | - |
| alu-v4_36 | 30 | 40 | - | - |
| 4gt13_92 | 17 | 24 | - | - |
| II. Chain, lower complexity | | | | |
| 4gt13_91 | 25 | 26 | 35 | - |
| 4gt5_76 | 22 | 26 | 51 | - |
| alu-v0_26 | 28 | 32 | - | - |
| alu-v3_35 | 24 | 26 | 34 | - |
| 4mod5-v1_24 | 29 | 31 | 44 | - |
| III. Connected, higher complexity | | | | |
| 4gt10-v1_81 | 37* | - | - | 39 |
| one-two-three-v1_99 | 52* | - | - | 45 |
| one-two-three-v0_98 | 47* | - | - | 61 |
| aj-e11_165 | 24 | - | - | 36 |
| alu-v2_32 | 30 | - | - | 41 |

Table 4: CNOT counts of circuits from imb_qx set [71] synthesized with CPFlow , SQUANDER , QSearch and a hybrid combination QFast+Qiskit+SQUANDER . Bars indicate either a failure to synthesize a circuit or the absence of data. Results of CPFlow were obtained via the ADAPTIVE routine with the following options: `min_num_cp_gates`=20, `max_num_cp_gates`=100, `num_samples`=1000, `max_evals`=100 except for the gate counts marked with an asterisk. The latter were obtained under a different option set: `min_num_cp_gates`=40, `max_num_cp_gates`=60, `num_samples`=2000, `max_evals`=100. Results of the other software packages are reproduced from [34].

20 to 100 for 100 evaluations with 1000 samples each. Results are reported in Table 4 (CZ counts marked by asterisks are explained below).

Results for group I, targeting synthesis of lower complexity circuits on the fully connected topology, show a significant compression achieved by CPFlow compared to SQUANDER , averaging to approximately 25%. In contrast, for circuits of similar complexity on the chain topology, group II, the difference between CPFlow and SQUANDER is less noticeable, averaging to a 10% additional compression. It would be interesting to understand the role of topology in either approach in more detail. The last group III consists of the circuits that the SQUANDER package failed to synthesize on its own, being unable to generate initial templates for compression [34]. In our view, the reason is likely to be rooted in the local minimums problem, which grows more acute with an increasing gate count. The authors of Ref. [34] proposed an interesting workaround to generate initial templates using other software packages (Qiskit + QFast ) and then further compress them with SQUANDER . The resulting gate counts are reported in the last column of Table 4. The last two circuits in this group were synthesized by CPFlow along with the circuits from groups I and II and turned out to have far lower complexity than SQUANDER results suggest. On the other hand, the first three circuits indeed proved to be the hardest to synthesize, and in fact CPFlow found only poor or no decompositions at all for these circuits with the original search options. This lead us to initiate a second ADAPTIVE optimization with a narrower gate range (`min_num_cp_gates`=40, `max_num_cp_gates`=60) and increased amount of samples (`num_samples`=2000). Eventually, acceptable and even apparently efficient decompositions were found by CPFlow yielding an average compression of 25%. Yet, it also became apparent that gate counts above 40 are a very challenging target for the algorithm, with most trials yielding no prospective circuits at all.

We need to stress that figures in Table 4 should not be taken as an accurate performance comparison between the algorithms (and hence we do not report the consumed resources and runtimes). First, the low-level implementation and the processor that we used are very different from those employed in Ref. [34]. Second, we relied on the results obtained in Ref. [34] to tune hyperparameters of CPFlow in advance. Precise comparison should also have a clear metric such as the maximum compression efficiency, faster runtime, scalability etc. Our primary objective was to minimize gate counts of synthesized circuits, the task that CPFlow addressed with a promising efficiently and within a relatively short time frame.

# 7 Summary and outlook

In this paper, we presented a new approach to the variational synthesis into CNOT plus 1q gate set. We identified the problem of local minimums as a crucial yet underappreciated obstacle that needs to be addressed. In the absence of an efficient way to avoid local minimums we have made an extensive exploration of the initial conditions space an integral part of our scheme. We also proposed to use parametric 2q gates as a way to unify the architecture search with the continuous optimization of 1q gates into a single coherent optimization and demonstrated its efficiency. A recent work based on a similar idea [34] also showed significant improvement over more standard discrete architecture search [27].

Another contribution of this work is a technique to promote approximate numerical circuits to exact decompositions. As an example, for the Toffoli gates studied in this paper we found that often efficient decompositions can be translated into circuits with angles being rational multiples of $\pi$ and verified to yield the exact synthesis. The technique is basically a post-processing of the numerical synthesis results and can be adopted by other approximate synthesis frameworks, so we believe it should be of broader interest.

While capable of generating many interesting results our method has its limitations. The first is rather fundamental and common to all similar schemes. Since the input circuit is represented by means of the corresponding unitary matrix, only small scale circuits that are easy to simulate classically can be addressed. This however does not preclude using variational synthesis to optimize smaller building blocks of large scale useful quantum algorithms [33]. Exploring this direction is an important and practically relevant avenue for future work.

Scaling up the variational approach within the classically accessible regime appears to be mostly limited by the circuit complexity rather than the qubit count, e.g. it seems to be more difficult to efficiently compile a 4q circuit that requires many 2q gates than it is to compile a 5q circuit that can be represented using only a small amount of 2q gates. Importantly, the challenges associated with the higher complexity are not caused only by the combinatorial growth of the possible architectures alone, but also by the proliferation of the local minimums in the loss landscape. Our empirical results (cf Fig. 6) indicate that already for 4q circuits in the range from 12 to 50 2q gates the probability of a successful continuous optimization is less than 0.1% even for a correctly chosen architecture. This probability depends on numerous factors including details of the optimization procedure, distribution of the initial conditions and parametrization of the loss landscape. Detailed understanding of these mechanisms may lead to a dramatic improvement in variational compilers' efficiency.

There are numerous further possibilities to enhance variational synthesis of high-complexity unitaries. For example, if the unitary originates from a known quantum circuit, it could be possible to take advantage of this information. Splitting the original circuit in parts, each having lower complexity, and synthesising them separately may lead to better results. Another proposal is to use the original circuit as the starting template for the variational compression [34]. A recent work [73] has shown that modifying template architectures on the go can help to reduce both global minimums and barren plateaus. It would be interesting to see if some of the circuits generated by CPFlow, which have efficient gate count but insufficient fidelity, can serve as a useful starting point for the "burrowing" procedure suggested in [73]. Eventually, the viability and resource allowance of the variational synthesis must be justified by the payoff if provides for useful applications.

## Acknowledgments

We would like to thank A. Nikolaeva, P. Rakyta, and Z. Zimborás for stimulating discussions, L. Madden and O. Lockwood for their considerate feedback on the draft, and V. Dubinin for the help with numerics. We also thank the two anonymous reviewers at Quantum, whose suggestions helped to improve and clarify this manuscript. This work was supported by the Russian Roadmap on Quantum Computing (the development of the algorithm and benchmarks in Sec. 6; Contract No. 868-1.3-15/15-2021, October 5, 2021). We thank the Priority 2030 program at the National University of Science and Technology "MISIS" under the project K1-2022-027

# A Circuit refinement

Quantum circuits that result from numerical optimization performed by CPFlow typically have many redundant 1q gates, see Fig. 13(a) for an example.We propose a simple refinement procedure that helps to find a simpler representation for such circuits.

First, we test if an angle $a$ can be set to 0 without affecting the loss function. For instance, in the state preparation problem the initial round of $R_Z$ gates has no effect when applied to the all-zero state and hence can be simply omitted.

Next, we check if there are cancellations between pairs of rotation gates (not necessarily adjacent). As a function of a single angle any parametrized quantum circuit has the following simple form

$$U(a) = U_0 \cos a + U_1 \sin a \tag{9}$$

with some unitary matrices $U_0, U_1$. In turn, as a func-

(a) Original circuit

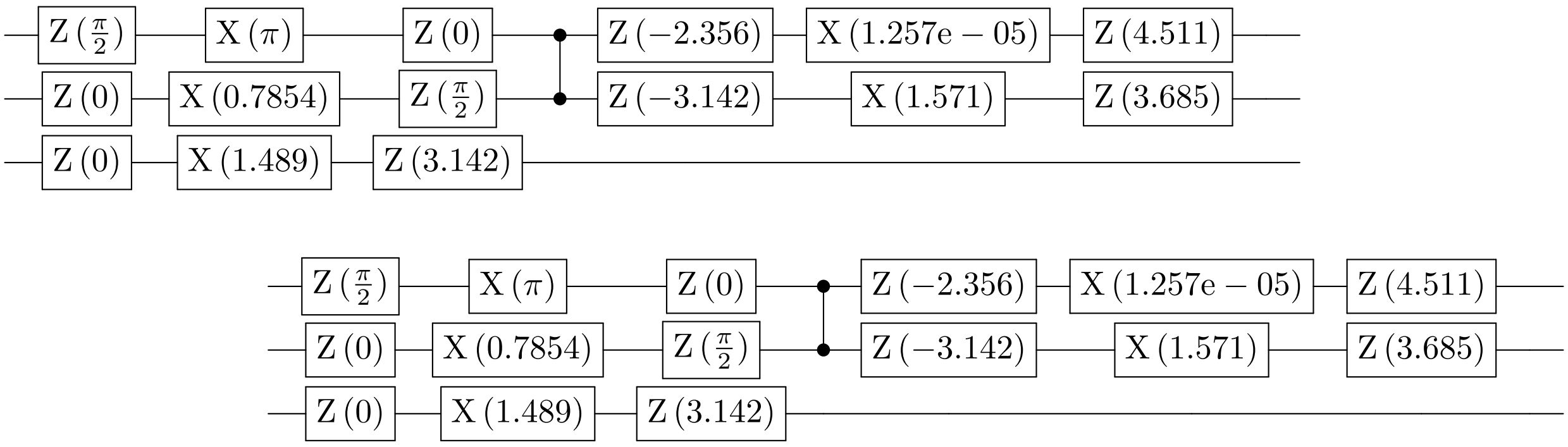

(b) Circuit with reduced angles

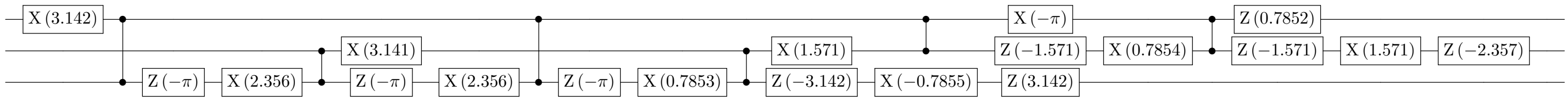

(c) Rationalized circuit

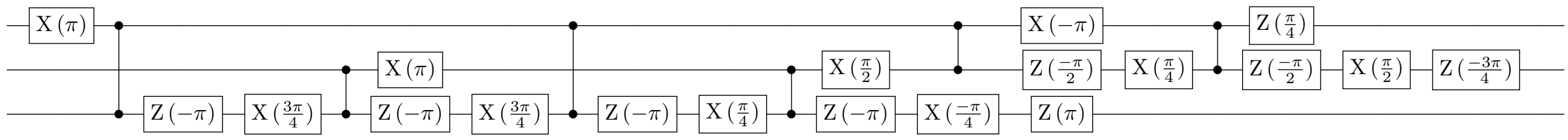

Figure 13: Refinement of the approximate decomposition of the 3q Toffoli gate into an exact result.

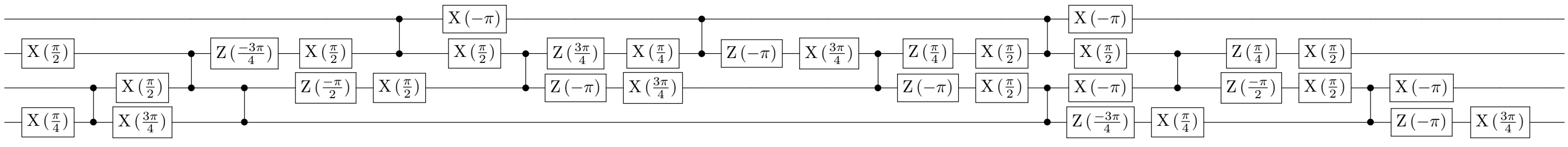

Figure 14: A decomposition of the relative phase 4q Toffoli gate on the chain topology with 11 CZ gates.

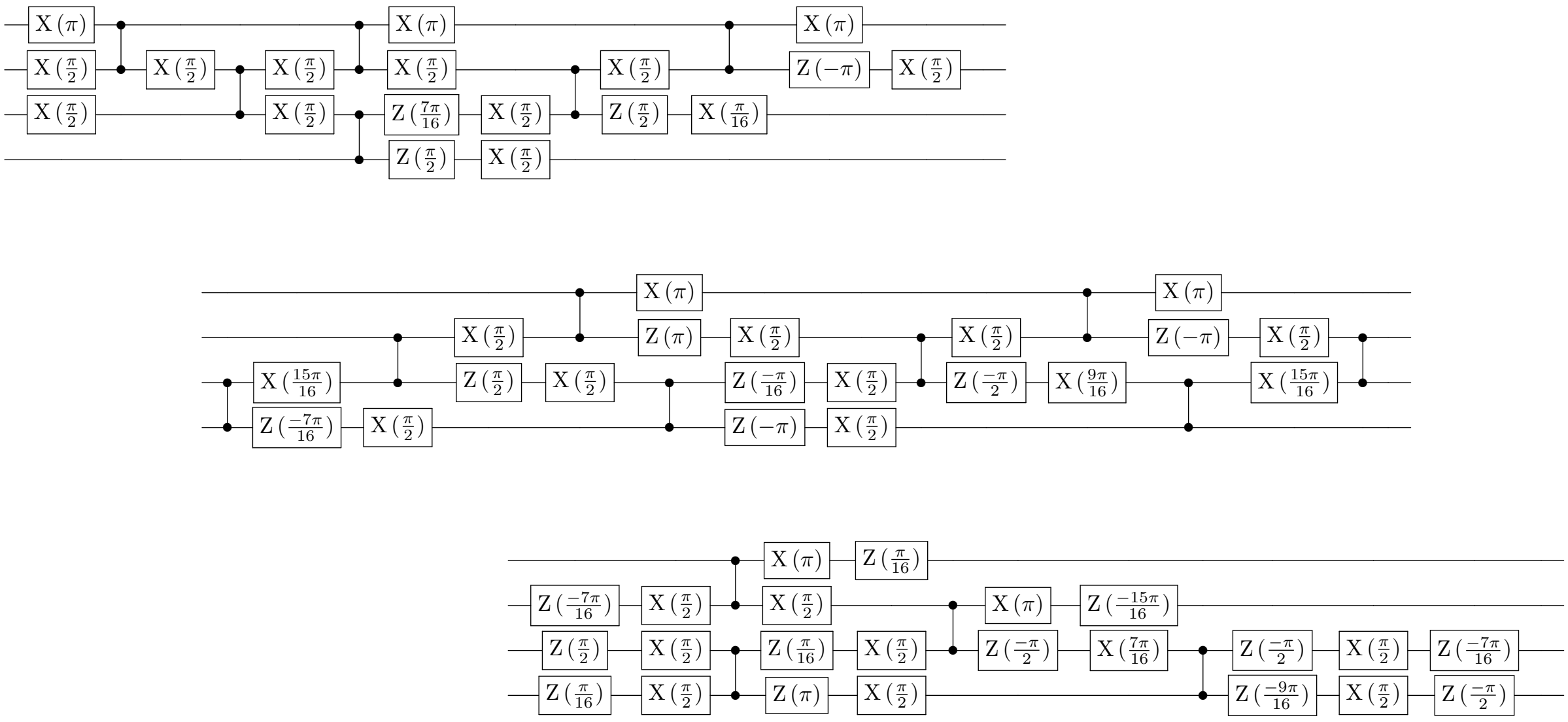

(8)

Figure 15: A decomposition of $\mathsf{C}^3\sqrt{\mathsf{X}}$ gate on the chain 4q topology with 18 CZ gates.

tion of two angles it can always be represented as

$$U(a_1,a_2) = U_{00}\cos a_1 \cos a_2 + U_{01}\cos a_1 \sin a_2 + U_{10}\sin a_1 \cos a_2 + U_{11}\sin a_1 a_2 \quad (10)$$

or, equivalently, as

$$\frac{U_{00}-U_{11}}{2}\cos(a_1+a_2) + \frac{U_{01}+U_{10}}{2}\sin(a_1+a_2) + \frac{U_{00}+U_{11}}{2}\cos(a_1-a_2) + \frac{U_{10}-U_{01}}{2}\sin(a_1-a_2)\ . \quad (11)$$

Typically, all terms in this expression are non-vanishing and different choices of $a_1$ and $a_2$ correspond to different unitaries (up to discrete redundancies). It may happen, however, that either the unitary $U$ itself or the loss function of interest $L(U)$ does not contain terms with $a_1 + a_2$ or $a_1 - a_2$. For instance, for the two consecutive $R_X$ rotations $R_X(a_1)R_X(a_2)$ only the terms that depend on the sum of angles are present in (11). Hence, for every pair of angles (it is usually sufficient to only consider angles of gates acting on the same qubit) we check if $L(U(0, a_2 \pm a_1)) = L(U(a_1,a_2))$ and if such pair is found, the first angle is set to 0 and the second angle is adjusted accordingly. Results of this step are illustrated at Fig. 13(b).

Finally, we check if the resulting angles of the 1q gates can be approximated by the rational multiples of $\pi$ without compromising the accuracy of the loss function $L(U)$ (in fact the accuracy is often improved at this step), see Fig. 13(c).

The steps outlined above are heuristic and do not always lead to expected results, but often work well in practice. All circuits reported in this paper were obtained automatically in this fashion. Heuristically we find that when decompositions are close to optimal, the refinement procedure works best. Possibly this can be attributed to the fact that extra gates allow for more complicated redundancies in the circuit that are not accounted for by our simple steps. Also, the procedure works best when the loss function is the most restrictive as in the compilation problem, when only the global phase of the unitary is not defined. When the loss function is more permissive, such as in state preparation, further steps usually need to be taken to eliminate all redundant gates.

# B 4q gates featuring in the decomposition of the 5q Toffoli gate

In Sec. 5.3 we constructed a decomposition of the 5q Toffoli gate on the chain topology with 48 CZ gates. This decomposition used efficient representations for the square root of the 4q Toffoli gate $\mathsf{C}^3\sqrt{\mathsf{X}}$ and a relative-phase 4q Toffoli gate on the chain 4q topology. Decomposition of $\mathsf{C}^3\sqrt{\mathsf{X}}$ with 18 CZ gates can be found using standard methodology described in Sec.5.2. The resulting circuit is depicted at Fig. 15. To find a relative-phase Toffoli gate one needs to use a non-standard loss function. By definition [67], $U$ is a relative phase Toffoli gate if $U = VD$, where $V$ is the unitary matrix of the Toffoli gate and $D$ is a diagonal unitary matrix. To construct the corresponding loss function we can use the fact that the sum $\sum_i |D_{ii}|^2$ for a unitary matrix $D$ has the maximum value $2^n$ iff $D$ is diagonal. Hence, the following loss reaches its minimum iff $U$ is a relative phase Toffoli gate

$$L(U) = 1 - \frac{\mathrm{Tr}\left|UV^{\dagger}\right|^2}{2^n} = 1 - \frac{\sum_{i,j}\left|U_{ij}V^*_{ji}\right|^2}{2^n} \quad (12)$$

With this loss function and standard parameter specifications for the 4q circuits used in this work CPFlow generated a relative phase 4q Toffoli gate on the chain topology with 11 CZ gates. The circuit is shown at Fig.14.

## References

[1] Peter W. Shor. "Polynomial-time algorithms for prime factorization and discrete logarithms on a quantum computer". SIAM Journal on Computing **26**, 1484–1509 (1997). arXiv:9508027.

[2] Lov K. Grover. "Quantum mechanics helps in searching for a needle in a haystack". Physical Review Letters **79**, 325–328 (1997).

[3] Aram W. Harrow, Avinatan Hassidim, and Seth Lloyd. "Quantum algorithm for linear systems of equations". Physical Review Letters **103**, 1–15 (2009). arXiv:0811.3171v3.

[4] A. K. Fedorov, N. Gisin, S. M. Beloussov, and A. I. Lvovsky. "Quantum computing at the quantum advantage threshold: a down-to-business review" (2022). arXiv:2203.17181.

[5] John Preskill. "Quantum computing in the NISQ era and beyond". Quantum **2**, 1–20 (2018). arXiv:1801.00862.

[6] Mohammadsadegh Khazali and Klaus Mølmer. "Fast multi-qubit gates by adiabatic evolution in interacting excited state manifolds" (2020). arXiv:2006.07035.

[7] Qiskit contributors (2023). code: https://doi.org/10.5281/zenodo.2573505.

[8] Seyon Sivarajah, Silas Dilkes, Alexander Cowtan, Will Simmons, Alec Edgington, and Ross Duncan. "t|ket: a retargetable compiler for NISQ devices". Quantum Science and Technology**6** (2021). arXiv:2003.10611v3.

[9] Adriano Barenco, Charles H. Bennett, Richard Cleve, David P. Divincenzo, Norman Margolus, Peter Shor, Tycho Sleator, John A. Smolin, and Harald Weinfurter. "Elementary gates for quantum computation". Physical Review A **52**, 3457–3467 (1995). arXiv:9503016.

[10] Y. Kharkov, A. Ivanova, E. Mikhantiev, and A. Kotelnikov. "Arline Benchmarks: Automated Benchmarking Platform for Quantum Compilers" (2022). arXiv:2202.14025.

[11] Panagiotis Kl Barkoutsos, Jerome F. Gonthier, Igor Sokolov, Nikolaj Moll, Gian Salis, Andreas Fuhrer, Marc Ganzhorn, Daniel J. Egger, Matthias Troyer, Antonio Mezzacapo, Stefan Filipp, and Ivano Tavernelli. "Quantum algorithms for electronic structure calculations: Particle-hole Hamiltonian and optimized wavefunction expansions". Physical Review A **98**, 022322 (2018). arXiv:1805.04340.

[12] Chufan Lyu, Victor Montenegro, and Abolfazl Bayat. "Accelerated variational algorithms for digital quantum simulation of many-body ground states". Quantum **4**, 324 (2020). arXiv:2006.09415v3.

[13] Sergey Bravyi, Alexander Kliesch, Robert Koenig, and Eugene Tang. "Obstacles to Variational Quantum Optimization from Symmetry Protection". Physical Review Letters **125**, 260505 (2020).

[14] Chufan Lyu, Xusheng Xu, Man-Hong Yung, and Abolfazl Bayat. "Symmetry enhanced variational quantum spin eigensolver". Quantum **7**, 899 (2023). arXiv:2203.02444.

[15] Lukasz Cincio, Kenneth Rudinger, Mohan Sarovar, and Patrick J. Coles. "Machine Learning of Noise-Resilient Quantum Circuits". PRX Quantum **2**, 010324 (2021). arXiv:2007.01210.

[16] Yuhan Huang, Qingyu Li, Xiaokai Hou, Rebing Wu, Man-Hong Yung, Abolfazl Bayat, and Xiaoting Wang. "Robust resource-efficient quantum variational ansatz through evolutionary algorithm". Physical Review A **105**, 052414 (2022). arXiv:2202.13714.

[17] Yuxuan Du, Tao Huang, Shan You, Min Hsiu Hsieh, and Dacheng Tao. "Quantum circuit architecture search for variational quantum algorithms". npj Quantum Information 2022 8:1 **8**, 1–8 (2022). arXiv:2010.10217.

[18] Hanrui Wang, Yongshan Ding, Jiaqi Gu, Yujun Lin, David Z. Pan, Frederic T. Chong, and Song Han. "QuantumNAS: Noise-Adaptive Search for Robust Quantum Circuits". Proceedings - International Symposium on High-Performance Computer Architecture **2022-April**, 692–708 (2021). arXiv:2107.10845.

[19] Sophia Fuhui Lin, Sara Sussman, Casey Duckering, Pranav S Mundada, Jonathan M Baker, Rohan S Kumar, Andrew A Houck, and Frederic T Chong. "Let Each Quantum Bit Choose Its Basis Gates" (2022). arXiv:2208.13380.

[20] D. P. Divincenzo and J. Smolin. "Results on two-bit gate design for quantum computers". Proceedings Workshop on Physics and Computation, PhysComp 1994Pages 14–23 (1994). arXiv:9409111.

[21] Matthew Amy, Dmitri Maslov, Michele Mosca, and Martin Roetteler. "A Meet-in-the-Middle Algorithm for Fast Synthesis of Depth-Optimal Quantum Circuits". IEEE Transactions on Computer-Aided Design of Integrated Circuits and Systems **32**, 818–830 (2013). arXiv:1206.0758.

[22] Harsha Nagarajan, Owen Lockwood, and Carleton Coffrin. "QuantumCircuitOpt: An Open-

source Framework for Provably Optimal Quantum Circuit Design". Proceedings of QCS 2021: 2nd International Workshop on Quantum Computing Software, Held in conjunction with SC 2021: The International Conference for High Performance Computing, Networking, Storage and AnalysisPages 55–63 (2021). arXiv:2111.11674.

[23] Yunseong Nam, Neil J. Ross, Yuan Su, Andrew M. Childs, and Dmitri Maslov. "Automated optimization of large quantum circuits with continuous parameters". npj Quantum Information**4** (2018). arXiv:1710.07345.

[24] Sumeet Khatri, Ryan LaRose, Alexander Poremba, Lukasz Cincio, Andrew T. Sornborger, and Patrick J. Coles. "Quantum-assisted quantum compiling". Quantum**3** (2019). arXiv:1807.00800.

[25] Harper R. Grimsley, Sophia E. Economou, Edwin Barnes, and Nicholas J. Mayhall. "An adaptive variational algorithm for exact molecular simulations on a quantum computer". Nature Communications 2019 10:1 **10**, 1–9 (2019). arXiv:1812.11173.

[26] Ho Lun Tang, V. O. Shkolnikov, George S. Barron, Harper R. Grimsley, Nicholas J. Mayhall, Edwin Barnes, and Sophia E. Economou. "Qubit-ADAPT-VQE: An Adaptive Algorithm for Constructing Hardware-Efficient Ansätze on a Quantum Processor". PRX Quantum **2**, 1–15 (2021). arXiv:1911.10205v2.

[27] Ethan Smith, Marc Grau Davis, Jeffrey Larson, E D Younis, Lindsay Bassman Oftelie, Wim Lavrijsen, Costin Iancu, Marc Grau Davis, Ed Younis, Bassman Lindsay, Wim Oftelie, and Costin Lavrijsen. "LEAP: Scaling Numerical Optimization Based Synthesis Using an Incremental Approach". ACM Transactions on Quantum Computing **4**, 1–23 (2023). arXiv:2106.11246.

[28] Richard Meister, Cica Gustiani, and Simon C. Benjamin. "Exploring ab initio machine synthesis of quantum circuits" (2022). arXiv:2206.11245.

[29] D. Chivilikhin, A. Samarin, V. Ulyantsev, I. Iorsh, A. R. Oganov, and O. Kyriienko. "MoG-VQE: Multiobjective genetic variational quantum eigensolver" (2020). arXiv:2007.04424.

[30] Lukas Franken, Bogdan Georgiev, Sascha Mücke, Moritz Wolter, Raoul Heese, Christian Bauckhage, and Nico Piatkowski. "Quantum Circuit Evolution on NISQ Devices" (2020). arXiv:2012.13453.

[31] Thomas Fösel, Murphy Yuezhen Niu, Florian Marquardt, and Li Li. "Quantum circuit optimization with deep reinforcement learning" (2021). arXiv:2103.07585.

[32] Mathias Weiden, John Kubiatowicz, Ed Younis, and Costin Iancu. "Ansatz Learning for Quantum Circuit Optimization". Bulletin of the American Physical Society (2023). url: https://meetings.aps.org/Meeting/MAR23/Session/G70.6.

[33] Ed Younis, Koushik Sen, Katherine Yelick, and Costin Iancu. "QFAST: Conflating Search and Numerical Optimization for Scalable Quantum Circuit Synthesis". Proceedings - 2021 IEEE International Conference on Quantum Computing and Engineering, QCE 2021Pages 232–243 (2021). arXiv:2103.07093.

[34] Péter Rakyta and Zoltán Zimborás. "Efficient quantum gate decomposition via adaptive circuit compression" (2022). arXiv:2203.04426.

[35] Bobak Toussi Kiani, Seth Lloyd, and Reevu Maity. "Learning Unitaries by Gradient Descent" (2020). arXiv:2001.11897.

[36] Liam Madden and Andrea Simonetto. "Best Approximate Quantum Compiling Problems". ACM Transactions on Quantum Computing **3**, 1–29 (2021). arXiv:2106.05649.

[37] Péter Rakyta and Zoltán Zimborás. "Approaching the theoretical limit in quantum gate decomposition" (2021). arXiv:2109.06770.

[38] Ken M. Nakanishi, Takahiko Satoh, and Synge Todo. "Quantum-gate decomposer" (2021). arXiv:2109.13223.

[39] Vivek V. Shende, Igor L. Markov, and Stephen S. Bullock. "Smaller two-qubit circuits for quantum communication and computation". Proceedings - Design, Automation and Test in Europe Conference and Exhibition **2**, 980–985 (2004).

[40] Vivek V. Shende, Stephen S. Bullock, and Igor L. Markov. "Synthesis of quantum-logic circuits". IEEE Transactions on Computer-Aided Design of Integrated Circuits and Systems **25**, 1000–1010 (2006). arXiv:0406176.

[41] Adi Botea, Akihiro Kishimoto, and Radu Marinescu. "On the complexity of quantum circuit compilation". Proceedings of the 11th International Symposium on Combinatorial Search, SoCS 2018Pages 138–142 (2018).

[42] Jarrod R. McClean, Jonathan Romero, Ryan Babbush, and Alán Aspuru-Guzik. "The theory of variational hybrid quantum-classical algorithms". New Journal of Physics **18**, 1–20 (2016). arXiv:1509.04279.

[43] Abhinav Kandala, Antonio Mezzacapo, Kristan Temme, Maika Takita, Markus Brink, Jerry M. Chow, and Jay M. Gambetta. "Hardware-efficient Variational Quantum Eigensolver for Small Molecules and Quantum Magnets". Nature **549**, 242–246 (2017). arXiv:1704.05018.

[44] Jarrod R. McClean, Sergio Boixo, Vadim N. Smelyanskiy, Ryan Babbush, and Hartmut Neven. "Barren plateaus in quantum neural network training landscapes". Nature Communications **9**, 1–7 (2018). arXiv:1803.11173.

[45] Lennart Bittel and Martin Kliesch. "Training Variational Quantum Algorithms Is NP-Hard".

Physical Review Letters **127**, 120502 (2021). arXiv:2101.07267.

[46] Martin Larocca, Nathan Ju, Diego García-Martín, Patrick J. Coles, and M. Cerezo. "Theory of overparametrization in quantum neural networks" (2021). arXiv:2109.11676.

[47] Xiaozhen Ge, Re-bing Wu, and Herschel Rabitz. "The Optimization Landscape of Hybrid Quantum-Classical Algorithms: from Quantum Control to NISQ Applications" (2022). arXiv:2201.07448.

[48] Eric R. Anschuetz. "Critical Points in Quantum Generative Models" (2021). arXiv:2109.06957.

[49] Eric R. Anschuetz and Bobak T. Kiani. "Quantum variational algorithms are swamped with traps". Nature Communications **13**, 7760 (2022). arXiv:2205.05786.

[50] Yiqing Zhou, E. Miles Stoudenmire, and Xavier Waintal. "What Limits the Simulation of Quantum Computers?". Physical Review X **10**, 1–14 (2020). arXiv:2002.07730.

[51] Andrea Skolik, Jarrod R. McClean, Masoud Mohseni, Patrick van der Smagt, and Martin Leib. "Layerwise learning for quantum neural networks". Quantum Machine Intelligence**3** (2020). arXiv:2006.14904v1.

[52] E. Campos, D. Rabinovich, V. Akshay, and J. Biamonte. "Training Saturation in Layerwise Quantum Approximate Optimisation". Physical Review A **104**, L030401 (2021). arXiv:2106.13814.

[53] David Wierichs, Christian Gogolin, and Michael Kastoryano. "Avoiding local minima in variational quantum eigensolvers with the natural gradient optimizer". Physical Review Research**2** (2020). arXiv:2004.14666.

[54] Javier Rivera-Dean, Patrick Huembeli, Antonio Acín, and Joseph Bowles. "Avoiding local minima in Variational Quantum Algorithms with Neural Networks" (2021). arXiv:2104.02955.

[55] Diederik P. Kingma and Jimmy Lei Ba. "Adam: A method for stochastic optimization". 3rd International Conference on Learning Representations, ICLR 2015 - Conference Track ProceedingsPages 1–15 (2015). arXiv:1412.6980.

[56] James Stokes, Josh Izaac, Nathan Killoran, and Giuseppe Carleo. "Quantum Natural Gradient". Quantum**4** (2020). arXiv:1909.02108.

[57] Tyson Jones and Simon C Benjamin. "Robust quantum compilation and circuit optimisation via energy minimisation". Quantum **6**, 628 (2022). arXiv:1811.03147.

[58] Owen Lockwood. "An Empirical Review of Optimization Techniques for Quantum Variational Circuits" (2022). arXiv:2202.01389.

[59] Robert Tibshirani. "Regression Shrinkage and Selection Via the Lasso". Journal of the Royal Statistical Society: Series B (Methodological) **58**, 267–288 (1996).

[60] David Donoho. "Donoho, d.l.: For most large underdetermined systems of linear equations the minimal l(1)-norm solution is also the sparsest solution. communications on pure and applied mathematics 59(6), 797-829". Communications on Pure and Applied Mathematics **59**, 797 – 829 (2006).

[61] Emmanuel Candes, Justin Romberg, and Terence Tao. "Robust Uncertainty Principles: Exact Signal Reconstruction from Highly Incomplete Frequency Information". IEEE Transactions on Information Theory **52**, 489–509 (2004). arXiv:0409186.

[62] Emmanuel J. Candès, Xiaodong Li, Yi Ma, and John Wright. "Robust principal component analysis?". Journal of the ACM **58**, 1–37 (2011).

[63] James Bergstra, Daniel Yamins, and David Cox. "Making a science of model search: Hyperparameter optimization in hundreds of dimensions for vision architectures". In Sanjoy Dasgupta and David McAllester, editors, Proceedings of the 30th International Conference on Machine Learning. Volume 28 of Proceedings of Machine Learning Research, pages 115–123. Atlanta, Georgia, USA (2013). PMLR. url: `https://proceedings.mlr.press/v28/bergstra13.html`.

[64] Nikita Nemkov, Ilia Luchnikov, Evgeniy Kiktenko, and Aleksey Fedorov. "cpflow". `https://github.com/idnm/cpflow` (2022).

[65] James Bradbury, Roy Frostig, Peter Hawkins, Matthew James Johnson, Chris Leary, Dougal Maclaurin, George Necula, Adam Paszke, Jake VanderPlas, Skye Wanderman-Milne, and Qiao Zhang (2018). url: `http://github.com/google/jax`.

[66] Guang Song and Andreas Klappenecker. "The simplified Toffoli gate implementation by Margolus is optimal" (2003). arXiv:quant-ph/0312225.

[67] Dmitri Maslov. "Advantages of using relative-phase Toffoli gates with an application to multiple control Toffoli optimization". Physical Review A **93**, 022311 (2016). arXiv:1508.03273.

[68] Vivek V. Shende and Igor L. Markov. "On the cnot-cost of toffoli gates". Quantum Information and Computation **9**, 461–486 (2009). arXiv:0803.2316.

[69] Norbert Schuch. "Implementation of quantum algorithms with josephson charge qubits". PhD thesis. Universität Regensburg. (2002). url: `https://epub.uni-regensburg.de/1511/`.

[70] Ken M. Nakanishi, Keisuke Fujii, and Synge Todo. "Sequential minimal optimization for quantum-classical hybrid algorithms". Physical Review Research **2**, 1–11 (2020). arXiv:1903.12166.

[71] Alwin Zulehner, Stefan Hillmich, Alexandru Paler, and Robert Wille. “ibm_qx_mapping”. https://github.com/iic-jku/ibm_qx_mapping (2017).

[72] Alwin Zulehner, Alexandru Paler, and Robert Wille. “An Efficient Methodology for Mapping Quantum Circuits to the IBM QX Architectures”. IEEE Transactions on Computer-Aided Design of Integrated Circuits and Systems **38**, 1226–1236 (2019). arXiv:1712.04722.

[73] Harper R. Grimsley, George S. Barron, Edwin Barnes, Sophia E. Economou, and Nicholas J. Mayhall. “Adaptive, problem-tailored variational quantum eigensolver mitigates rough parameter landscapes and barren plateaus”. npj Quantum Information **9**, 19 (2023). arXiv:2204.07179.